\documentclass[aps,prl,showpacs,twocolumn,floats,superscriptaddress]
{revtex4}
\usepackage{graphicx}
\usepackage{bm}
\usepackage{color}
\makeatletter
\def\lsim{\mathrel{\mathpalette\gl@align<}}
\def\gsim{\mathrel{\mathpalette\gl@align>}}
\def\gl@align#1#2{\lower.6ex\vbox
{\baselineskip\z@skip\lineskip\z@
\ialign{$\m@th#1\hfil##\hfil$\crcr#2\crcr\sim\crcr}}}

\makeatother
\usepackage{amsmath}
\usepackage{color}
\usepackage{graphicx}
\usepackage{amsfonts}
\usepackage{mathrsfs}
\usepackage{amsmath}
\begin{document}

\title{Bosons with incommensurate potential and spin-orbit coupling}

\author{Sayak Ray}
\affiliation{Indian Institute of Science Education and Research,
Kolkata, Mohanpur, Nadia 741246, India}

\author{Bhaskar Mukherjee}
\affiliation{Theoretical Physics Department, Indian Association for
the Cultivation of Science, Jadavpur, Kolkata-700032, India.}

\author{S. Sinha}
\affiliation{Indian Institute of Science Education and Research,
Kolkata, Mohanpur, Nadia 741246, India}

\author{K. Sengupta}
\affiliation{Theoretical Physics Department, Indian Association for
the Cultivation of Science, Jadavpur, Kolkata-700032, India.}

\date{\today}

\begin{abstract}

We chart out the phase diagram of ultracold `spin-half' bosons in a
one-dimensional optical lattice in the presence of Aubry-Andr\'e
(AA) potential and with spin-orbit (SO) and Raman couplings
investigating the transition from superfluid (SF) to localized
phases and the existence of density wave phase for nearest-neighbor
interaction (NNI). We show that the presence of SO coupling and AA
potential leads to a novel spin-split momentum distribution of the
bosons in the localized phase near the boundary with the SF phase,
which can act as a signature of such a transition. We also obtain
the level statistics of the bosons in the superfluid phase with
finite NNI and demonstrate its change from Gaussian Unitary Ensemble (GUE) to
Gaussian Orthogonal Ensemble (GOE) as a function
of the Raman coupling. We discuss experiments which can test our
theory.

\end{abstract}


\maketitle
The study of localization phenomena in correlated systems has
regained a new interest recently in the context of many-body
localization (MBL) \cite{mbl1,mbl2}. Ultracold atoms in optical lattices,
which act as emulators of strongly correlated model Hamiltonians \cite{boserev1},
can serve as test beds for such phenomena \cite{boserev2}. In this
context, systems with quasiperiodic potentials, which have posed
several interesting theoretical challenges over may decades
\cite{quasiref1,quasiref2,quasiref3}, turn out to be
particularly relevant. A model Hamiltonian describing such a
quasiperiodic system is the well-known Aubry-Andr\'e (AA) model
\cite{aapaper}, which, unlike the Andreson model, exhibits
localization transition in 1D \cite{aapaper,aapaper1}. This property of the AA model has
generated an impetus to study MBL \cite{quasirev1,mblaa2}. Moreover, experimental
realization of the AA model in bichromatic optical lattice has led
to observation of localization of both light\cite{loclight1} and
ultracold matter wave \cite{inguscio,kushref1}.

In recent past, extensive research on the Bose-Hubbard (BH) model
using ultracold bosonic atoms in optical lattices paved the way for
studying the effect of interactions on localization phenomenon
leading to possible glassy phases
\cite{inguscio1,demarco,bhref1,Roth}. In addition, intense
theoretical studies has also been carried out on the BH model in the
presence of Abelian and non-Abelian gauge fields; such gauge fields
have been experimentally realized in atom-laser systems
\cite{ab1,nonab1}. Such systems allow for observation of several
exciting phenomena \cite{abph1,abph2, nonabph1}; most interestingly,
they enable us to study strongly interacting bosons in the presence
of tunable spin-orbit (SO) coupling
\cite{sorefs1,sorefs2,sorefs3,sorefs4}. The realization of the AA
model in bichromatic lattice and the creation of SO interactions for
ultracold bosons therefore provides an unique opportunity to study
localization phenomenon induced by the AA potential in presence of
tunable SO interactions.

In this work, we study a two-species Bose-Hubbard model coupled by
Raman frequency $\Omega$, in the presence of an AA potential and a
SO coupling and show that such a system leads to several novel
features which appear only in the presence of both the AA potential
and the SO coupling. The central results of our study are as
follows. First, we chart out the phase diagram of 1D ultracold
bosons in an optical lattice and demonstrate the existence of
density wave (DW), superfluid (SF), and localized phases and study
the transition between these phases. Second, we show that for
sufficiently high $\Omega$, the bosons in the presence of both the
AA potential and the SO coupling exhibits a spin-split momentum
distribution in the localized phase, near the boundary with the SF
phase, irrespective of the strength of their interaction. Such a
splitting can therefore serve as a signature of this transition. We
note that this spin splitting does not occur in the absence of
either the AA potential or the SO coupling. Third, we study the
level statistics of the bosons in the strongly interacting regime,
where the presence of AA potential and Raman coupling $\Omega$
between the spins play a crucial role in changing the spectral
statistics between different universality classes of random matrix
theory (RMT). Apart from poissonian level spacing distribution in
the localized regime, we find that the level statistics change
continually from GUE ($\Omega=0$) to GOE as a function of $\Omega$.
We identify the additional symmetry at the $\Omega=0$ point which is
behind this change. Finally, we discuss experiments which can test
our theory.

The Hamiltonian of the bosons in a bi-chromatic 1D lattice with AA
potential and SO coupling is given by
\begin{eqnarray}
\hat{H} &=& -t\sum_{l,\sigma}
\left(\hat{b}_{l,\sigma}^{\dagger}e^{iq\hat{\sigma}_z}
\hat{b}_{l+1,\sigma}+h.c.\right) + \frac{1}{2}\sum _{l, l^{\prime}}
\mathcal{V}_{l,l^{\prime}}\hat{n}_l \hat{n}_{l^{\prime}} \nonumber\\
&& +\lambda \sum_{l,\sigma}\cos(2\pi \beta l)\hat{n}_{l,\sigma} +
\Omega \sum_{l,\sigma}\hat{b}_{l,\sigma}^{\dagger} \hat{b}_{l,\bar
\sigma} \label{ham1}
\end{eqnarray}
where, $\hat{b}_{l,\sigma}^{\dagger}$ and $\hat{n}_{l
\sigma}=\hat{b}_{l,\sigma}^{\dagger} \hat{b}_{l,\sigma} $ are the
creation and the density operator of the bosons of (pseudo)spin
$\sigma$ at the lattice site $l$, $\hat{n}_l = \sum _{\sigma} \hat{n}_{l
\sigma}$, $\bar \sigma = \downarrow
(\uparrow)$ for $\sigma= \uparrow (\downarrow)$, $t$ is the hopping
strength, $q$ is the SO coupling strength, $\Omega$ is the Raman
frequency and $\lambda$ denotes the strength of the quasiperiodic
potential. In the rest of the paper we consider nearest neighbor and
on-site interactions with coupling strengths: $V =
\mathcal{V}_{l,l+1}$ and $U = \mathcal{V}_{l,l}$ respectively; in
what follows, we shall scale all energies in unit of $t$.

{\it Non-interacting limit}: We first look into the the
non-interacting bosons by setting $U = V = 0$ in Eq.\ \ref{ham1}.
For $\lambda=0$, this reduces to a pure SO coupled bosonic system
with single particle spectrum
\begin{equation}
E_k^{\pm} = -2\cos k \cos q \pm 2\left[\sin ^2 k \sin ^2 q + \Omega
^2/4\right]^{1/2} \label{spspec}
\end{equation}
and the eigenstates are given by, $\psi _k^{\pm} = e^{ikx} (
\cos (\theta_{k} -\pi/4\mp \pi/4), \sin (\theta_{k} -\pi/4 \mp
\pi/4))^T$, where $\cos \theta_{k} = [1/2 + [4 + \Omega
^2/(\sin ^2k \sin ^2q)]^{-1/2}]^{1/2}$.
We find that there exists a critical Raman coupling given by,
\begin{eqnarray}
\Omega _c &=& 2 \sin q \tan q
\end{eqnarray}
below which the ground state is doubly degenerate associated with
the finite momenta $k_0 = \pm \cos ^{-1}[\cos q(1 + \Omega ^2/(4\sin
^2 q))^{1/2}]$. The doubly degenerate ground states are related by
$\psi_{-k} = \hat{\sigma}_z \hat{\mathcal T} \psi_{k}$, where $\hat{\mathcal T} = -i\hat{\sigma} _y
\hat{\mathcal{C}}$ is the time reversal symmetry (TRS) operator and
$\hat{\mathcal{C}}$ is the complex conjugation operator.

Next we turn on $\lambda$ keeping $U=V=0$. For $q =\Omega =0$, above
Hamiltonian is reduced to a two component AA model which undergoes a
localization transition above a critical coupling strength
$\lambda_{c} =2$. For two extreme regimes $q\ne 0, \Omega=0$ (pure
SO coupling) and $q=0, \Omega \ne 0$ (strong Raman coupling) the
single particle Hamiltonian preserves the self duality at
$\lambda_{c} =2$ and all states are localized above $\lambda_c$. To
study localization transition in the intermediate regime with $q\ne
0, \Omega \ne 0$, we numerically diagonalize the single particle
Hamiltonian to obtain the ground state and the excitation spectrum.
Since the duality does not hold in this regime a mobility edge
appears and energy dependent localization occurs for eigenstates \cite{aaso}. We
focus on the localization transition of the ground state in presence
of SO interaction and the variation of the critical disorder
strength $\lambda_c$ on Raman coupling. We locate the change from the
SF to the localized phase in two ways. First, we measure the
superfluid fraction(SFF) by applying a phase twist \cite{Fisher}
$\theta \ll \pi$ at the boundary. In the presence of such a twist
$t_{l,l+1} \rightarrow te^{i\theta/N_s}$ (Eq.\ \ref{ham1}). The SFF
can then be computed as \cite{Roth},
\begin{equation}
f_{s} = N_s^2(E[\theta] -E[0])/(N_p\theta^2), \label{sff}
\end{equation}
where $E[\theta]$ is the ground state energy in presence of twist,
$N_s$($N_p$) is the number of sites(particles).
\begin{figure}[ht]
\centering
\includegraphics[width=4.1cm,height=3.2cm]{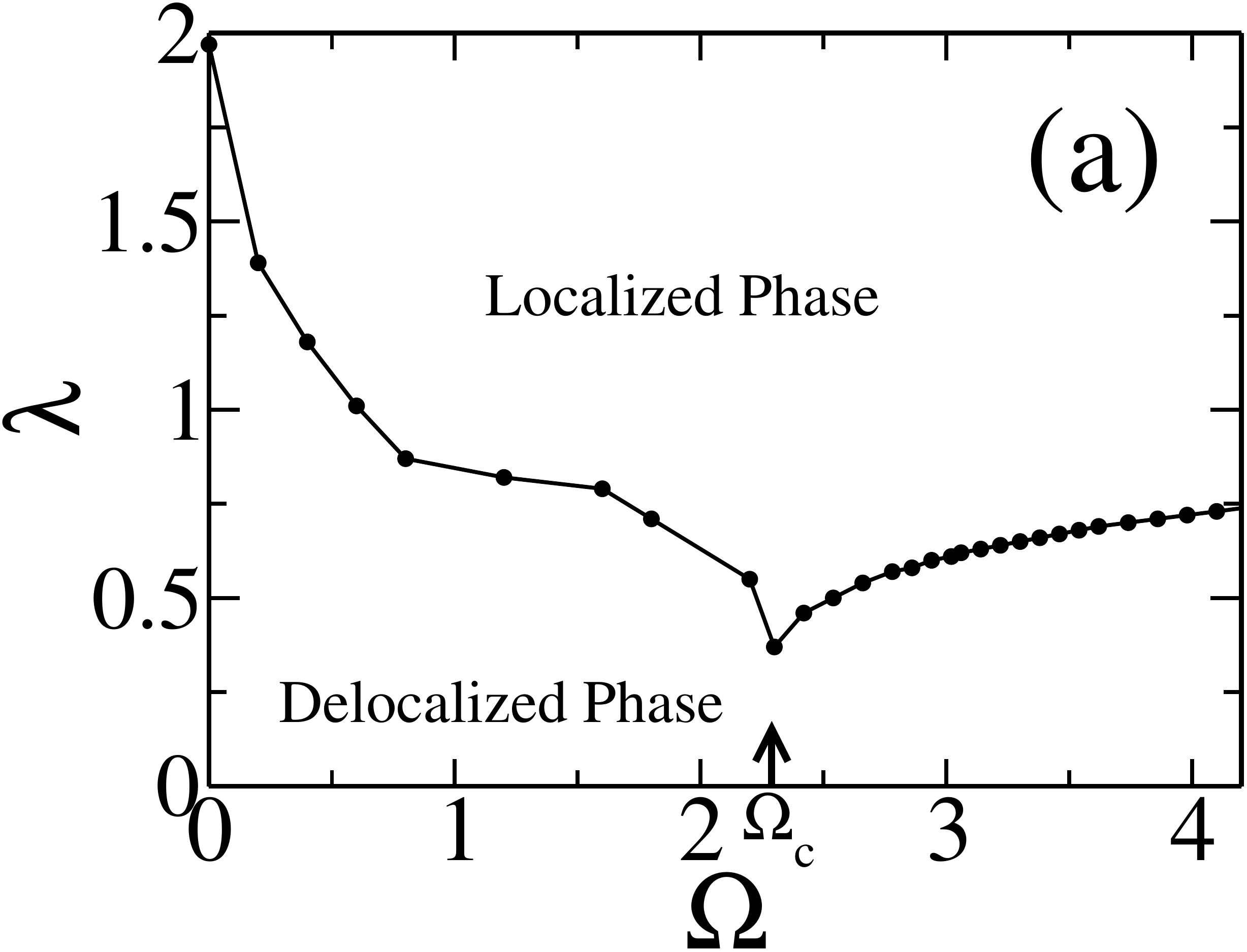}
\includegraphics[width=4.1cm,height=3.2cm]{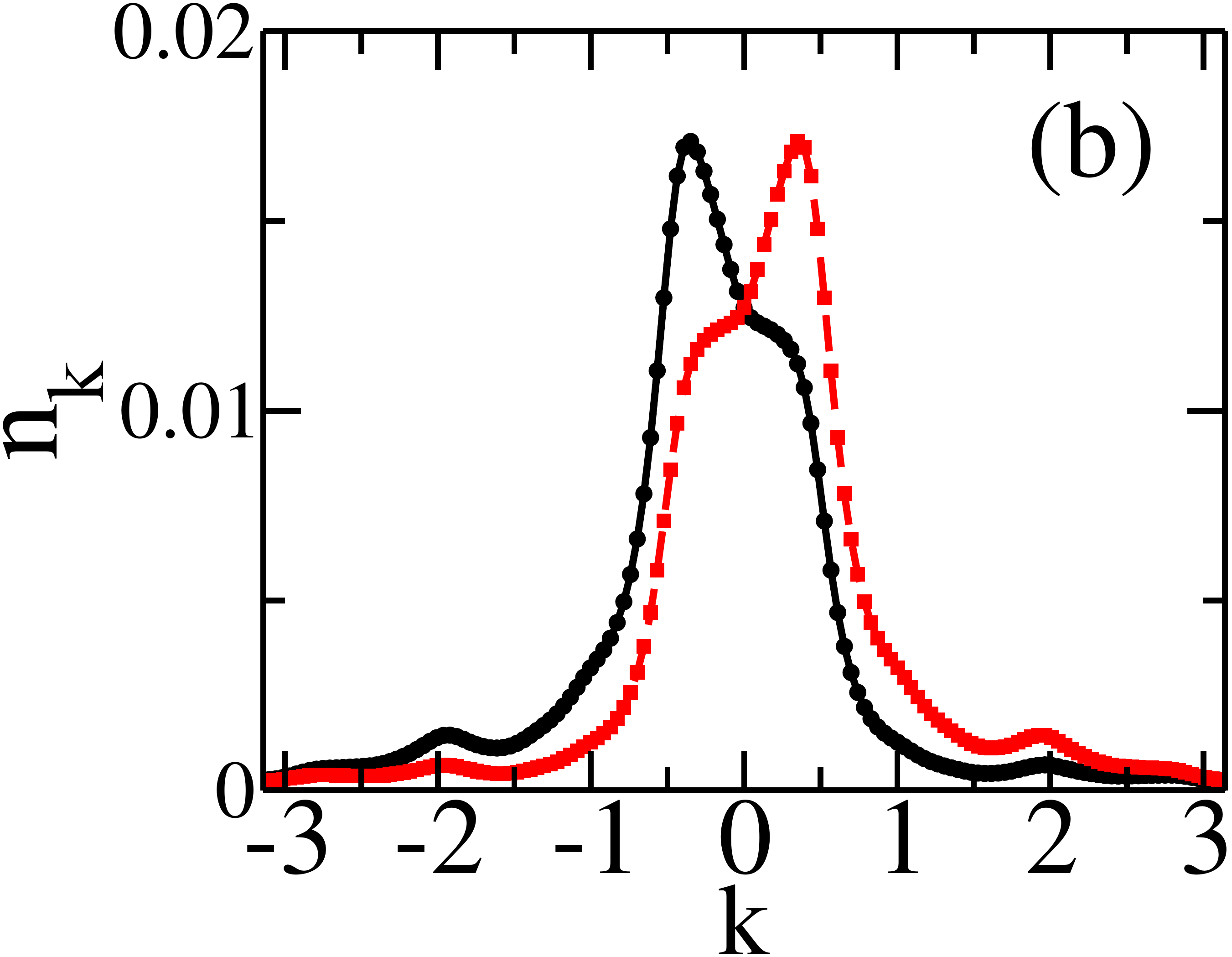}
\includegraphics[width=4.1cm,height=3cm]{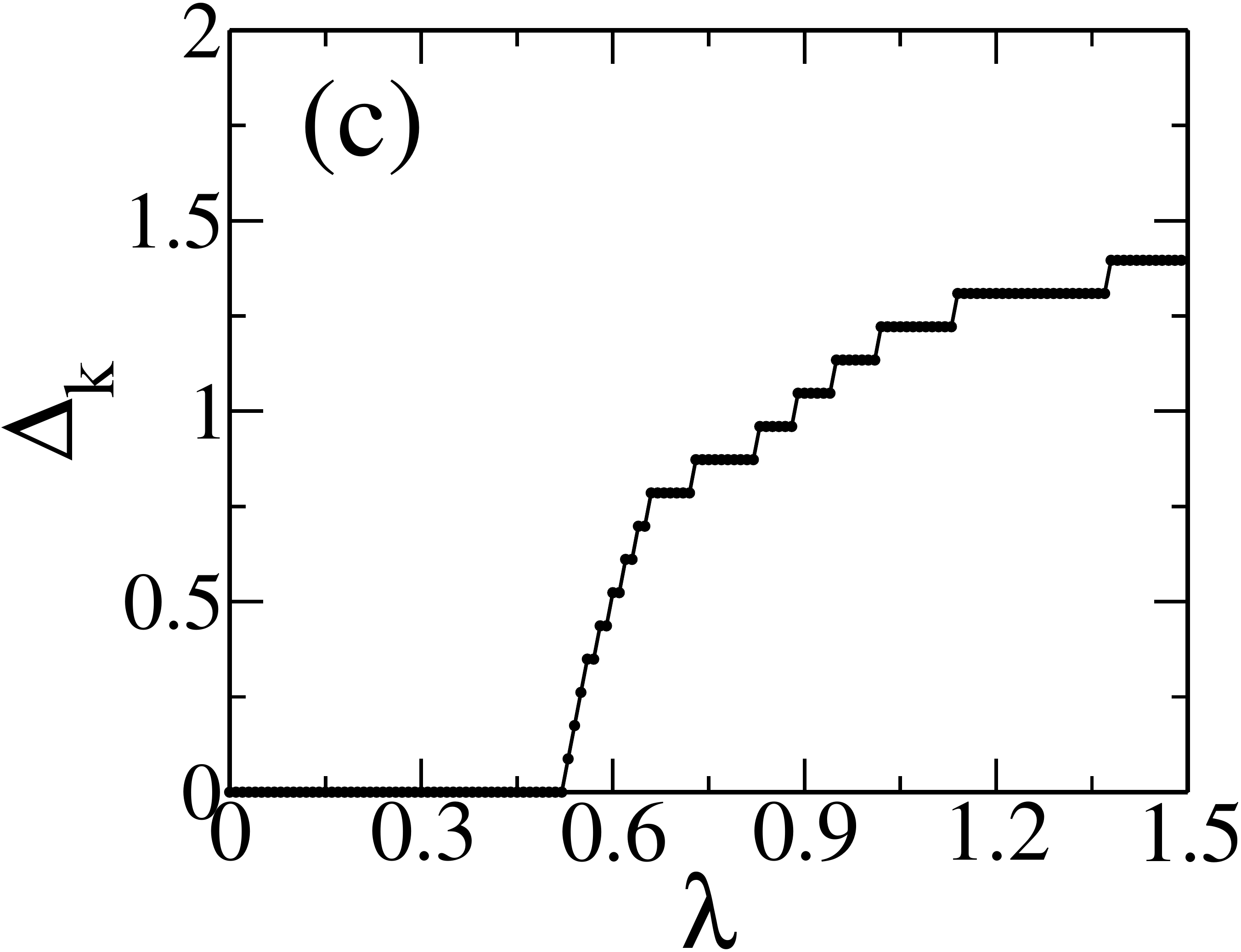}
\includegraphics[width=4.1cm,height=3cm]{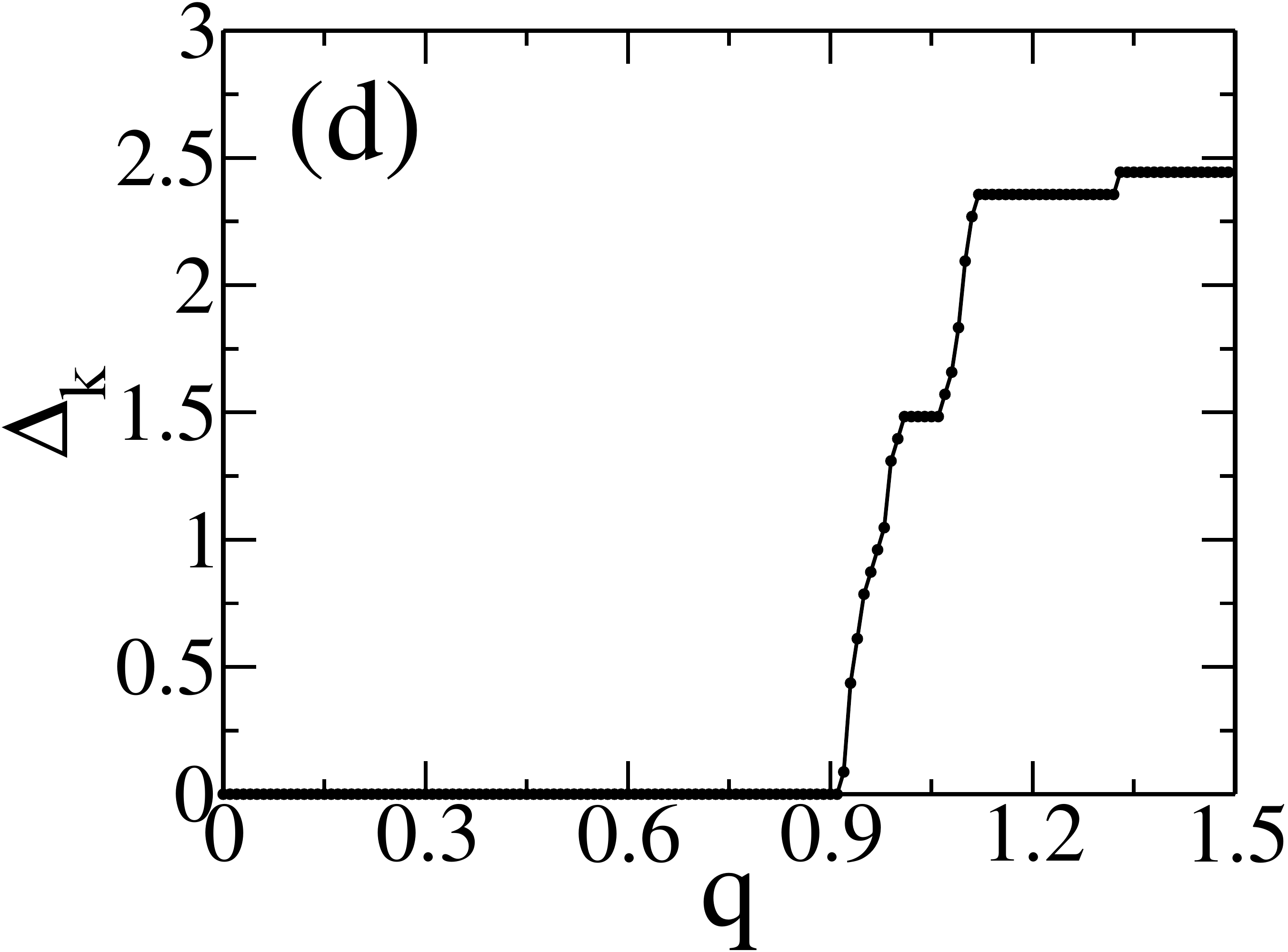}
\caption{(a) Phase diagram is shown in the $\Omega - \lambda$ plane.
(b) Momentum distribution in the delocalized phase is shown for $\Omega = 2.5$,
$\lambda=0.65$. Solid(dashed) curves correspond to the up(down) spin. (c-d) $\Delta_k$ as a
function of $\lambda$ for $q=0.3\pi$ and as a function of $q$ for $\lambda=0.65$ is shown.
For both the plots we set $\Omega = 2.5$.} \label{fig1}
\end{figure}
The second measure of localization is the inverse participation
ratio (IPR) of the ground state wavefunction defined as
\begin{equation}
I = \sum _l (|\psi _{l,\uparrow}|^2 + |\psi _{l,\downarrow}|^2)^2
\end{equation}
where $|\psi _{l,\sigma}|^2$ is the boson density of spin $\sigma$
at site $l$. As expected, we find that SFF decreases and the IPR
increases with increasing $\lambda$ around the transition.

The phase diagram obtained from these computations is shown in Fig.\
\ref{fig1}(a) in $\lambda-\Omega$ plane for $q=0.3 \pi$. We note that
$\lambda_c$ decreases from its self-dual value $\sim 2$
for $q \ne 0$ and shows a dip at $\Omega_c$, which
demarcates the delocalized phase in two regime. Below $\Omega_c$ the
degeneracy of the ground state is lifted by the quasi-periodic
potential; however the ground state has a net momentum and $s_z$
polarization. For $\Omega
> \Omega_c$, the ground state wavefunction is spin-polarized along
$\hat x$ and has vanishing net momentum.  The behavior of
$\lambda_c$ with $\Omega$ can be understood from the enhancement of
effective mass of bosons in the lower branch $m^{\ast} =
\partial ^2 E_k^{-}/\partial k^2|_{k = k_{0}}$ due to the combined
effect of SO and Raman couplings. This in turn reduces the effective
hopping strength $t_{\rm eff} = t/m^{\ast}$ of underlying AA model
for which the critical strength for localization transition can be
estimated as $\lambda_{c} \sim 2 t_{\rm eff} \sim 2/m^{\ast}$
\cite{supp1}. We note that the idea of $m^{\ast}$ also
quantitatively explains the the variation of SFF with $\Omega$ for
$\lambda =0$ and that SFF decreases and the IPR increases with
increasing $\lambda$ around the transition as expected \cite{supp1}.

\begin{figure}[ht]
\centering
\includegraphics[width=4.1cm,height=3.2cm]{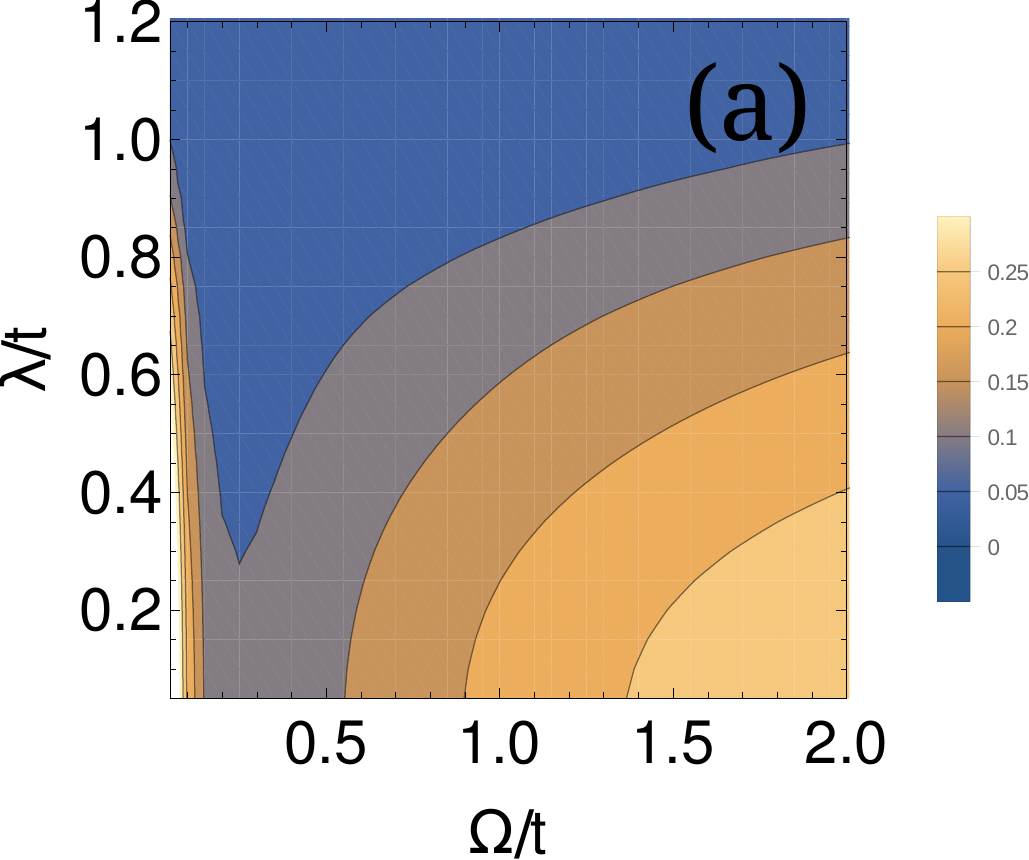}
\includegraphics[width=4.1cm,height=3.2cm]{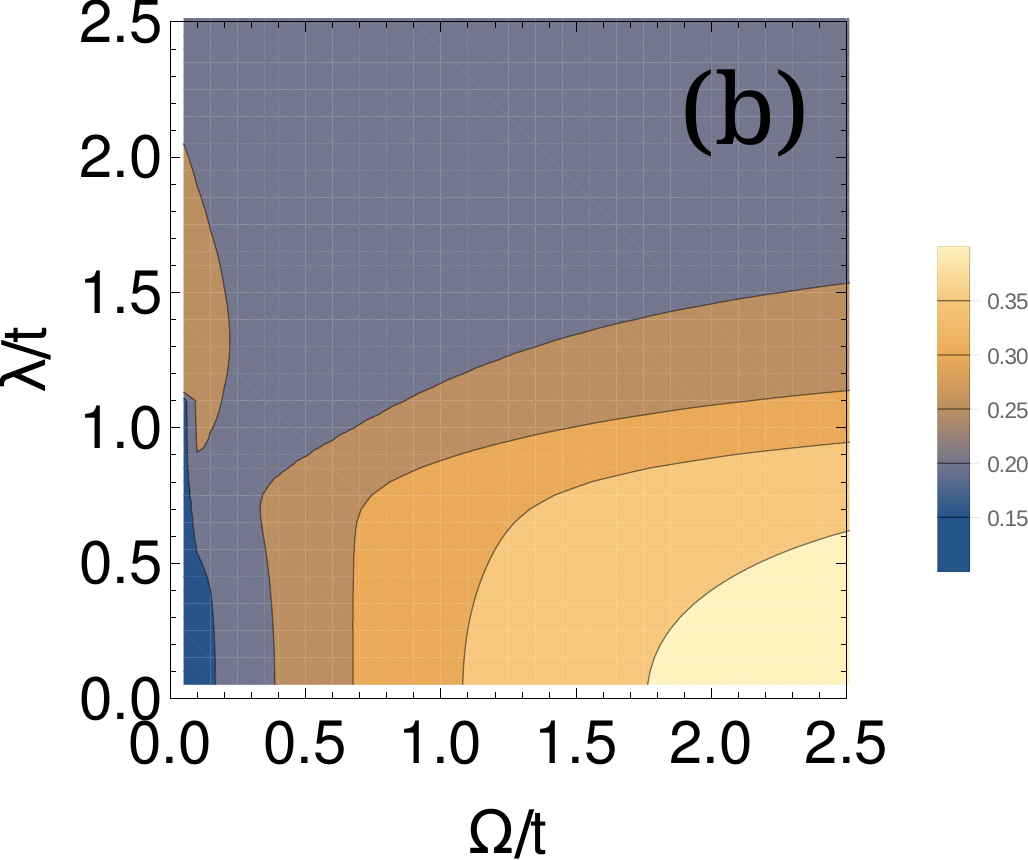}
\includegraphics[width=4.1cm,height=3cm]{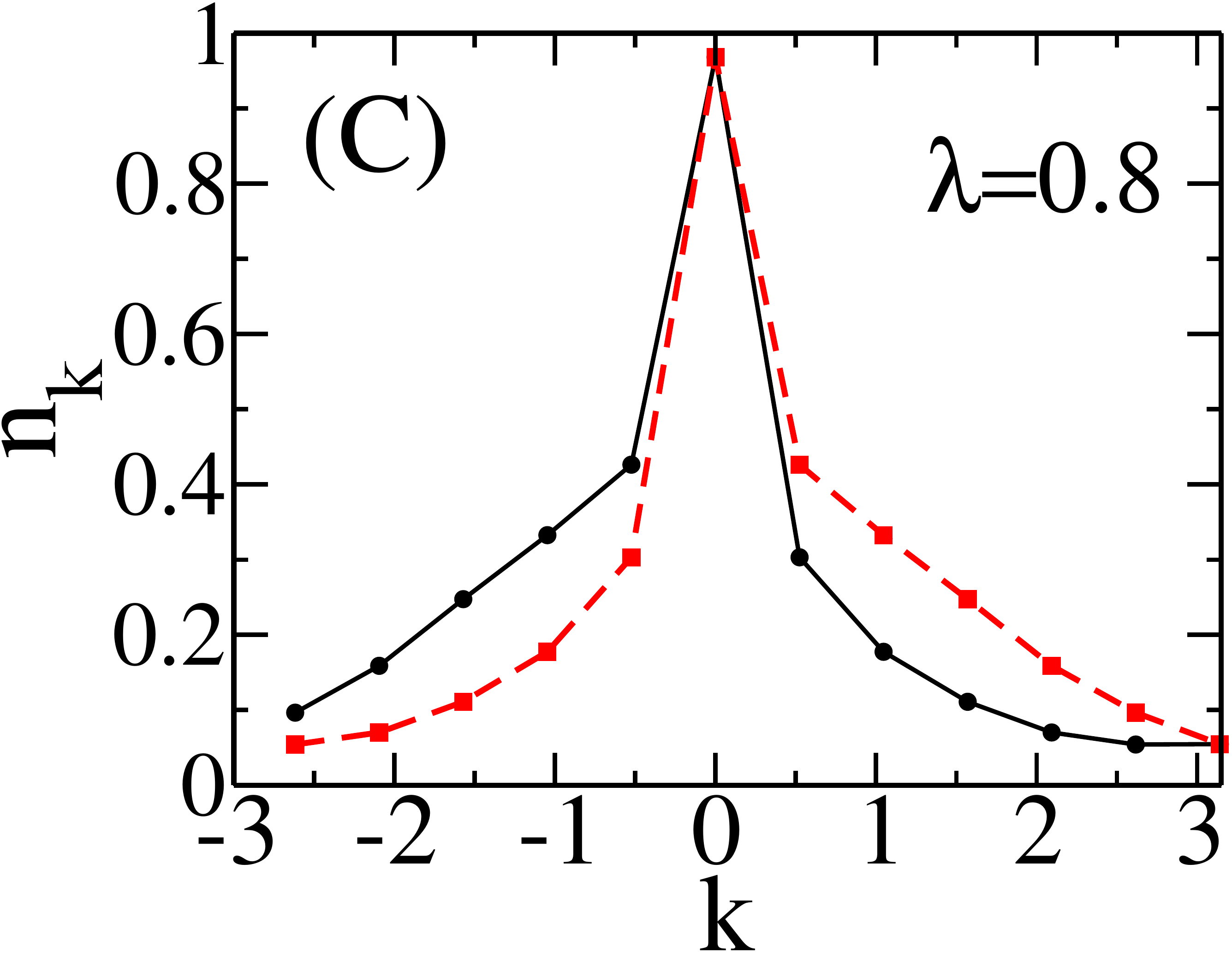}
\includegraphics[width=4.1cm,height=3cm]{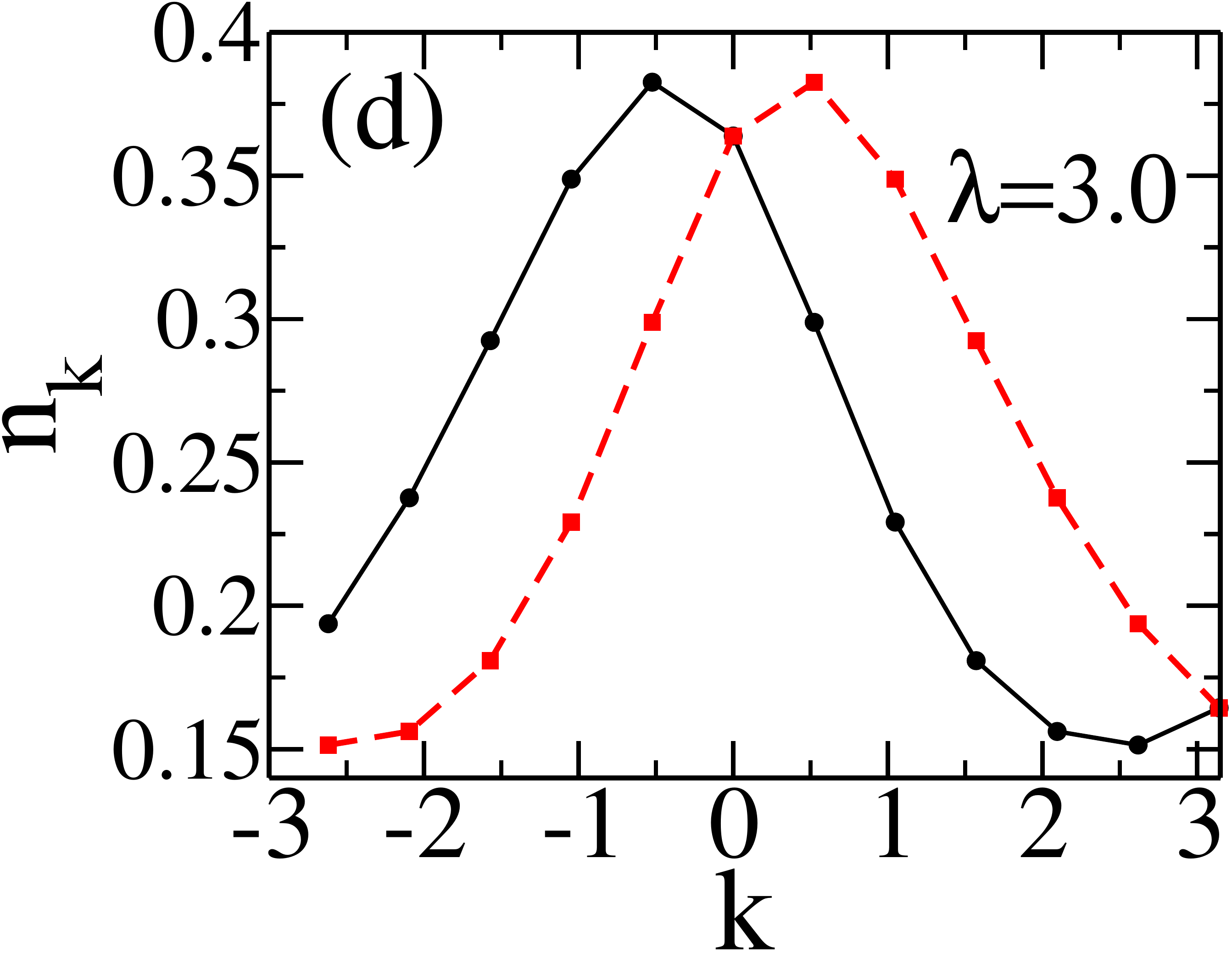}
\caption{(a)-(b) : Color plot of the SFF $f_s$ and
the BCF $f_c$ have been shown in the $\Omega - \lambda$
plane for $q = 0.3\pi$ and no. of sites $N_s = 12$ at
half filling. (c)-(d) : Momentum distribution of spin
up(down) particles are shown for $\Omega = 2.5$ by solid(dashed)
lines. See the text for details.} \label{fig2}
\end{figure}

To elucidate the role of the AA potential and the SO coupling in the
transition, we compute the spin-resolved momentum distribution
defined as
\begin{equation}
n_{k \sigma} = \sum_{l,l^{\prime}} \exp\{i k (l-l^{\prime})\} \langle c_{l \sigma}^{\dagger}
c_{l^{\prime} \sigma} \rangle/N_s \label{mdist}
\end{equation}
where $k=2\pi m/N_s$, with $m\in [-N_s/2,(N_s-1)/2]$. For $\Omega
> \Omega_c$, both $n_{k \uparrow}$ and $n_{k \downarrow}$ is peaked at $k=0$
in the delocalized phase as seen for standard superfluids. In
contrast, in the localized regime near the transition, $n_{k
\sigma}$ becomes spin dependent and is peaked at $k=k_{\sigma}^{\rm
max} \ne 0$ preserving the symmetry $n_{\uparrow}(k) =
n_{\downarrow}(-k)$ (see Fig.\ \ref{fig1}(b)). The splitting of
these peaks are given by $\Delta_{k^{\rm max}_{\sigma}} =
k_{\uparrow}^{\rm max}-k_{\downarrow}^{\rm max} \sim q$ leading to
the conclusion that the split in $n_{k \sigma}$ arises from a finite
SO coupling. As shown in Fig.\ \ref{fig1}(c), (d), $\Delta_{k^{\rm
max}_{\sigma}}$ vanishes for either $q=0$ or $\lambda=0$; this shows
the necessity of both the AA potential and the SO coupling for the
peak splitting. We find that this splitting can be qualitatively
understood from a variational wavefunction calculation and is
associated with spin dephasing showing a fluctuation of relative phase
between two spin components of the ground state wavefunction
\cite{supp1}.

{\it Hardcore limit}: To explore the effect of interaction on this
phenomenon we now set $U \to \infty$, keeping $V=0$. This limit
facilities computation by imposing the constraint $n_{l} \le
1$ at each site and allows us to perform exact diagonalization
within a restricted Hilbert space of three states per site. We
restrict our calculation to half-filled HC bosons, $\sum_l n_{l} =N_s/2$
so that we are always in the SF phase for
$\lambda=V=0$. In addition to SFF we also compute the boson
condensate fraction (BCF) since BCF and SFF are quite different for
strongly interacting bosons and are important for characterizing the
localized phases. We construct the one-body density matrix from the
ground state $|\psi_0\rangle$: $\rho(l,\sigma;l^{\prime},\sigma ^{\prime}) =
\langle \psi
_0|\hat{b}_{l^{\prime},\sigma^{\prime}}^{\dagger}\hat{b}_{l,\sigma}|\psi
_0\rangle$\cite{legg1}; the largest eigenvalue $N_c$ of which gives
the BCF $f_c = N_c/{\rm Tr}(\hat{\rho})$.

\begin{figure}[ht]
\centering
\includegraphics[width=4.1cm,height=3.2cm]{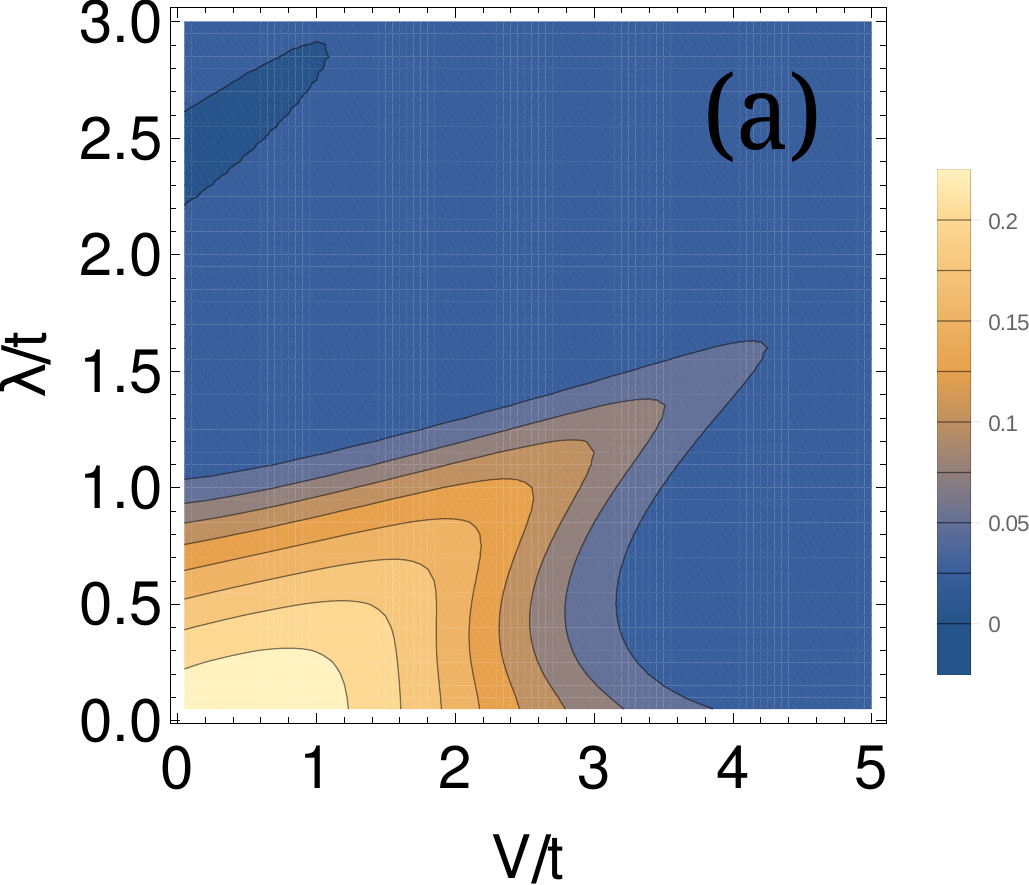}
\includegraphics[width=4.1cm,height=3.2cm]{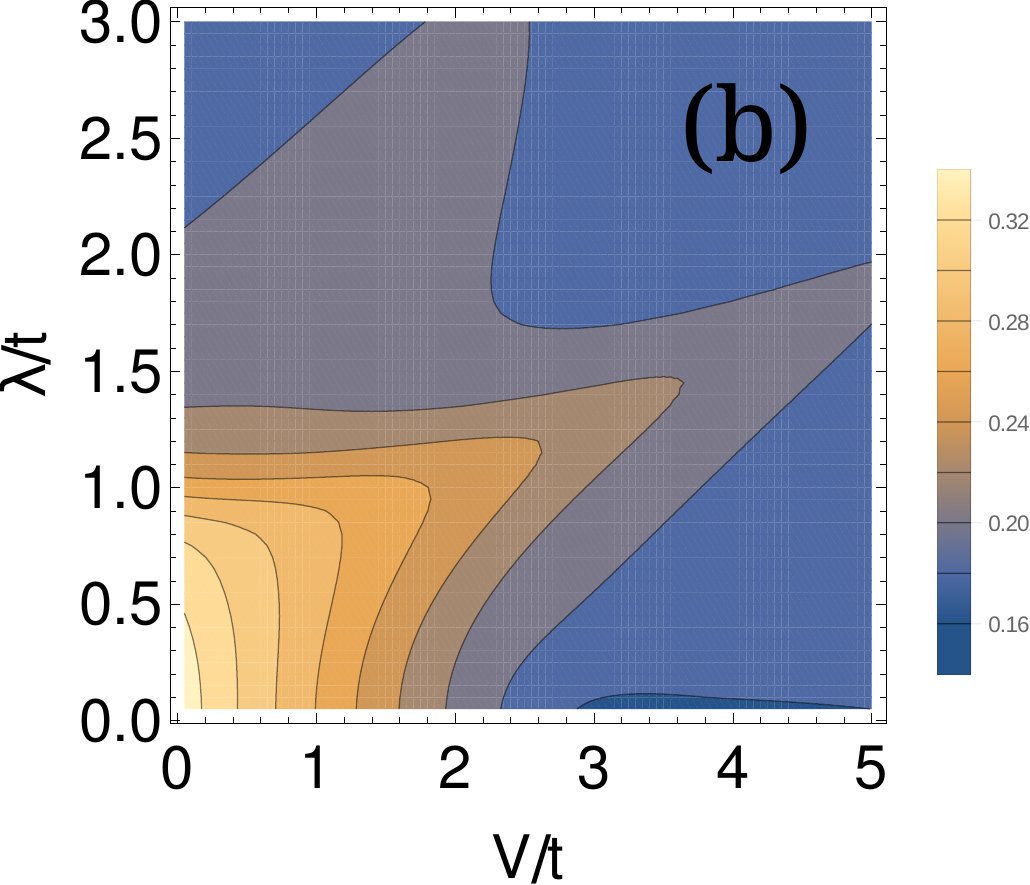}
\includegraphics[width=4.1cm,height=3cm]{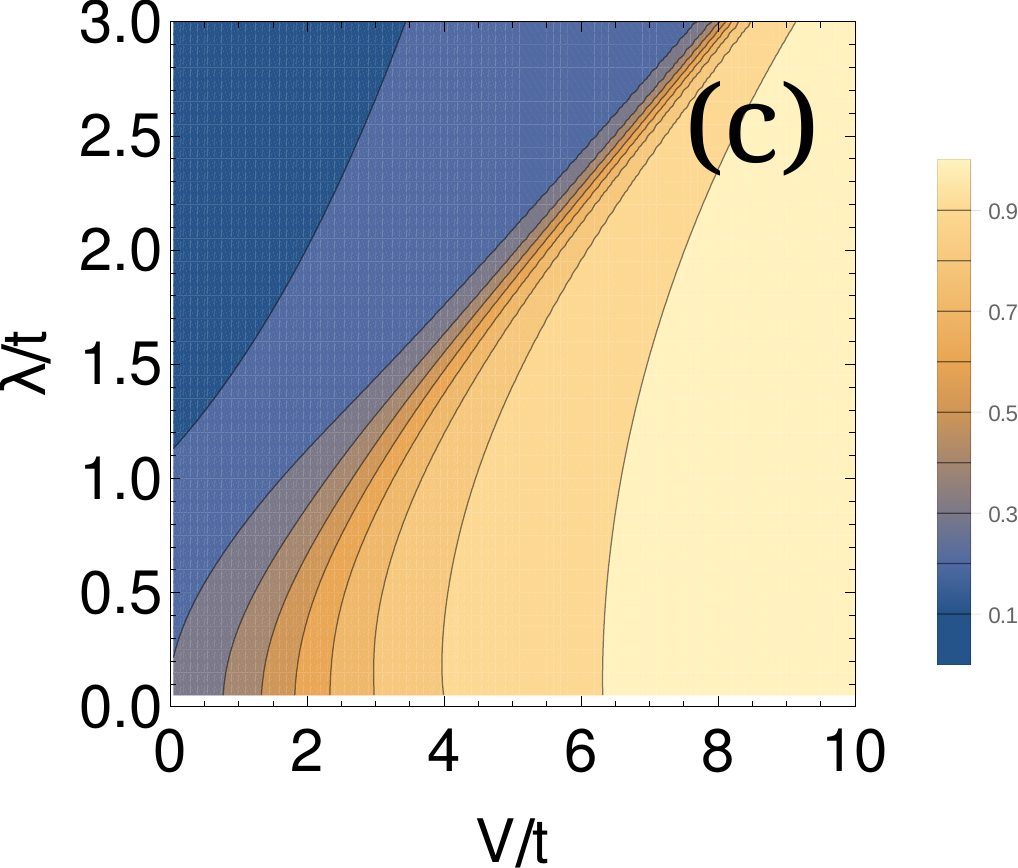}
\includegraphics[width=3.7cm,height=3cm]{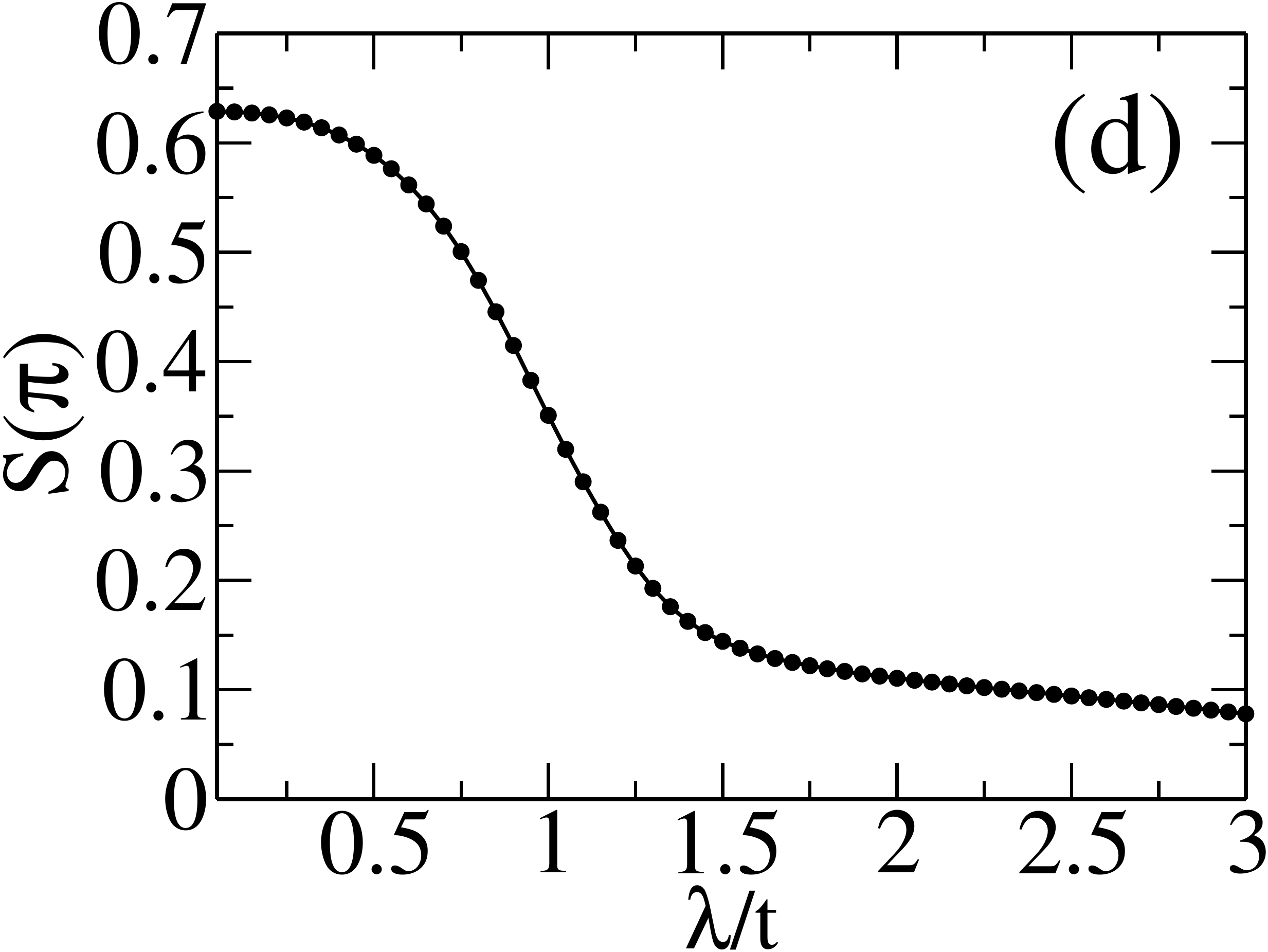}
\caption{Color plot of the (a) superfluid density, (b) condensate
fraction and (c) $S(\pi)$ for bosons at half filling with $N_s = 12$
as a function of $V/t$ and $\lambda/t$ for $\Omega = 1.5$ and $q=0.3
\pi$. (d) $S(\pi)$ as a function of $\lambda/t$ for $V/t=2.5$. See
the text for details.} \label{fig3}
\end{figure}

A plot of $f_s$ and $f_c$ in the $\Omega - \lambda$ plane for a
fixed $q$ is shown in Fig.\ \ref{fig2}(a,b). These plots clearly
indicate a regime for $\lambda > \lambda_c$ where $f_s$ vanishes but
$f_c$ remains finite indicating localized phase of the bosons.
Although in finite system there is no transition, the behavior of
$\lambda_c$ obtained from SFF is similar to that for non-interacting
bosons; however $\Omega_c$ shifts to a lower value. Near this
boundary, particularly for $\Omega \leq \Omega_c$, there is clear
indication of Bose-glass (BG) phase with $f_{s} =0$ and $f_{c}\ne
0$.

Finally, we compute the $n_{k \sigma}$ of the hardcore bosons. As
shown in Figs.\ \ref{fig2}(c),(d), the splitting of the spin momentum peak
occurs in the localized phase and survives in the hardcore limit. We
have checked that $n_{k \sigma}$ are peaked at $k = 0$ in the
delocalized regime and at $k_{\sigma}^{\rm max}$ in the localized
regime near the transition. Thus we find that the shift in $n_{k
\sigma}$ due to presence of a finite $q$ survives in the presence of
strong on-site interaction. Similar conclusions can be drawn for
weakly interacting bosons for which $U/t \ll 1$\cite{supp1}.

{\it Phase diagram at finite $V$}: Next we turn on a finite $V$ for
the hardcore bosons and obtain the phase diagram by computing SFF
and BCF as a function of $V/t$ and $\lambda/t$ for a fixed $q$ and
$\Omega$ (see Fig.\ \ref{fig3}(a,b)). For small $V$ we find that an
increase of $\lambda$ leads to a depletion of superfluid density
keeping the condensate fraction finite indicating a finite-size
crossover from a SF to a localized phase. Similarly for a fixed
small $\lambda$, an increase in $V$ leads to an analogous depletion
of superfluid density; this indicates the onset of the DW phase with
broken translational symmetry. For $\lambda = \Omega =0$,
Eq.\ref{ham1} reduces to the well studied XXZ model which exhibits
the SF to DW transition at $V/t =1$ \cite{XXZ}. Similarly for the
SF-DW transition at small $\lambda$, a first order transition is
expected since the DW state breaks translational invariance while
the SF states breaks $U(1)$ gauge symmetry.
\begin{figure}[ht]
\centering
\includegraphics[width=4.1cm,height=3.2cm]{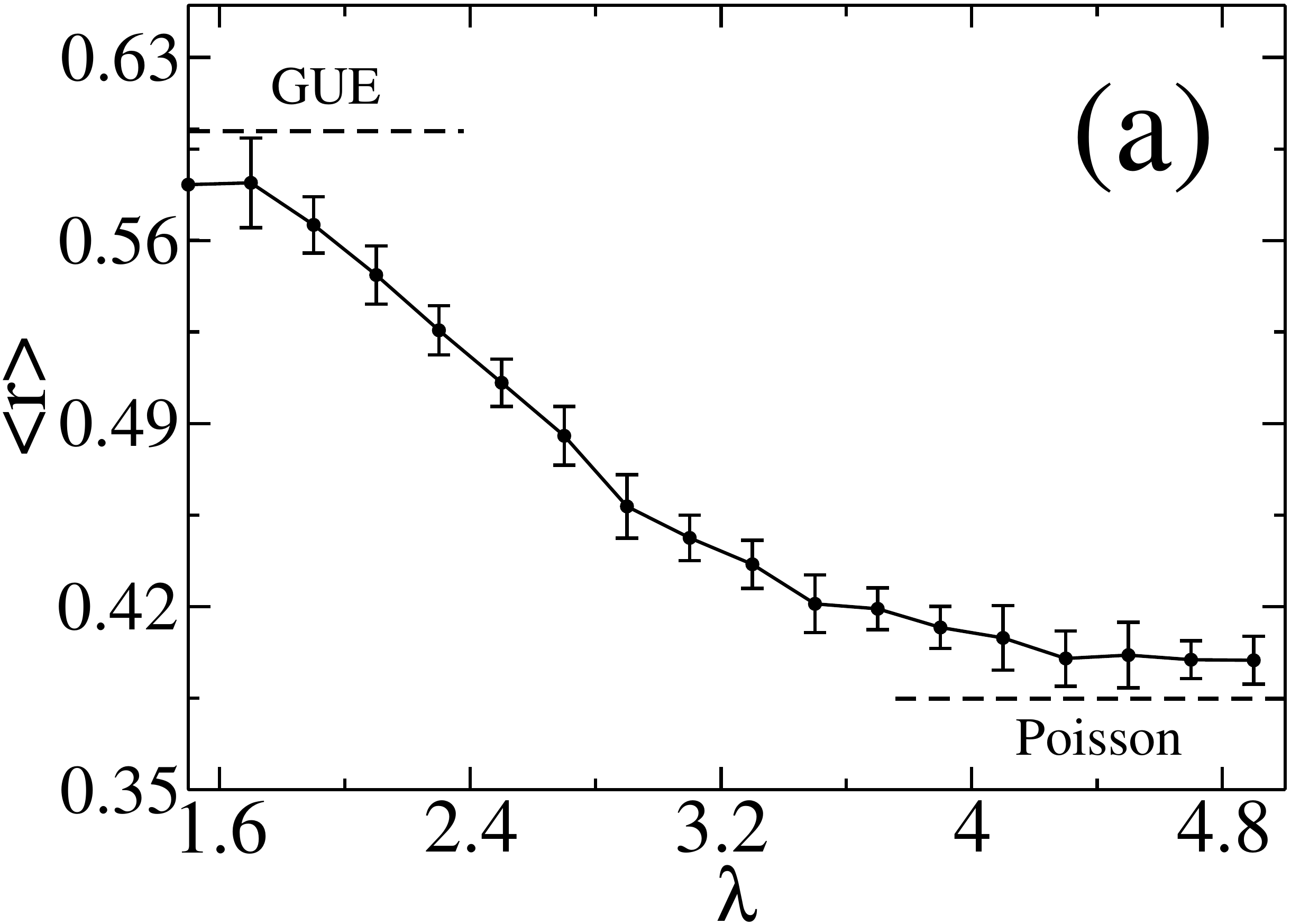}
\includegraphics[width=4.1cm,height=3.2cm]{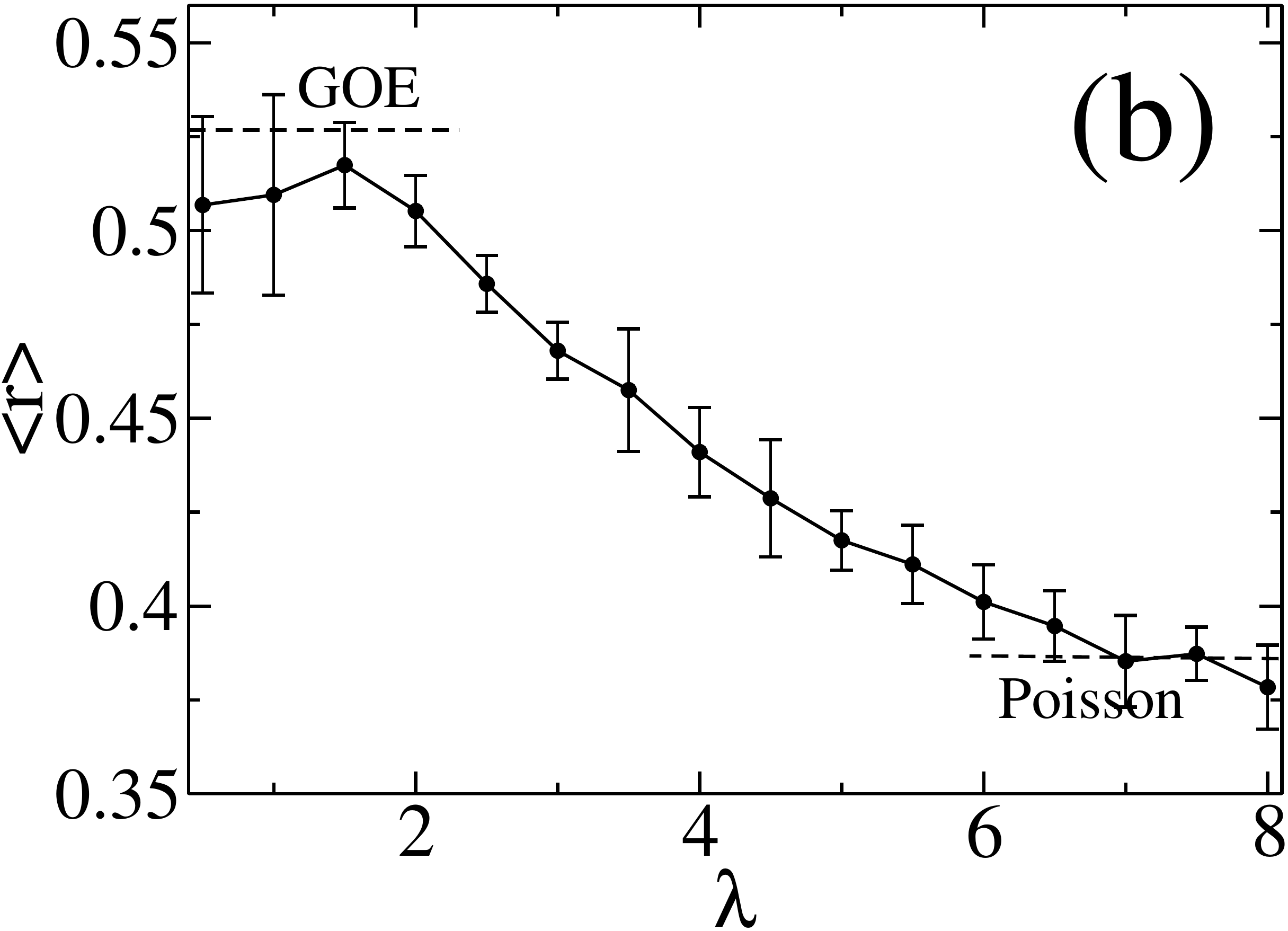}
\includegraphics[width=4.1cm,height=3.2cm]{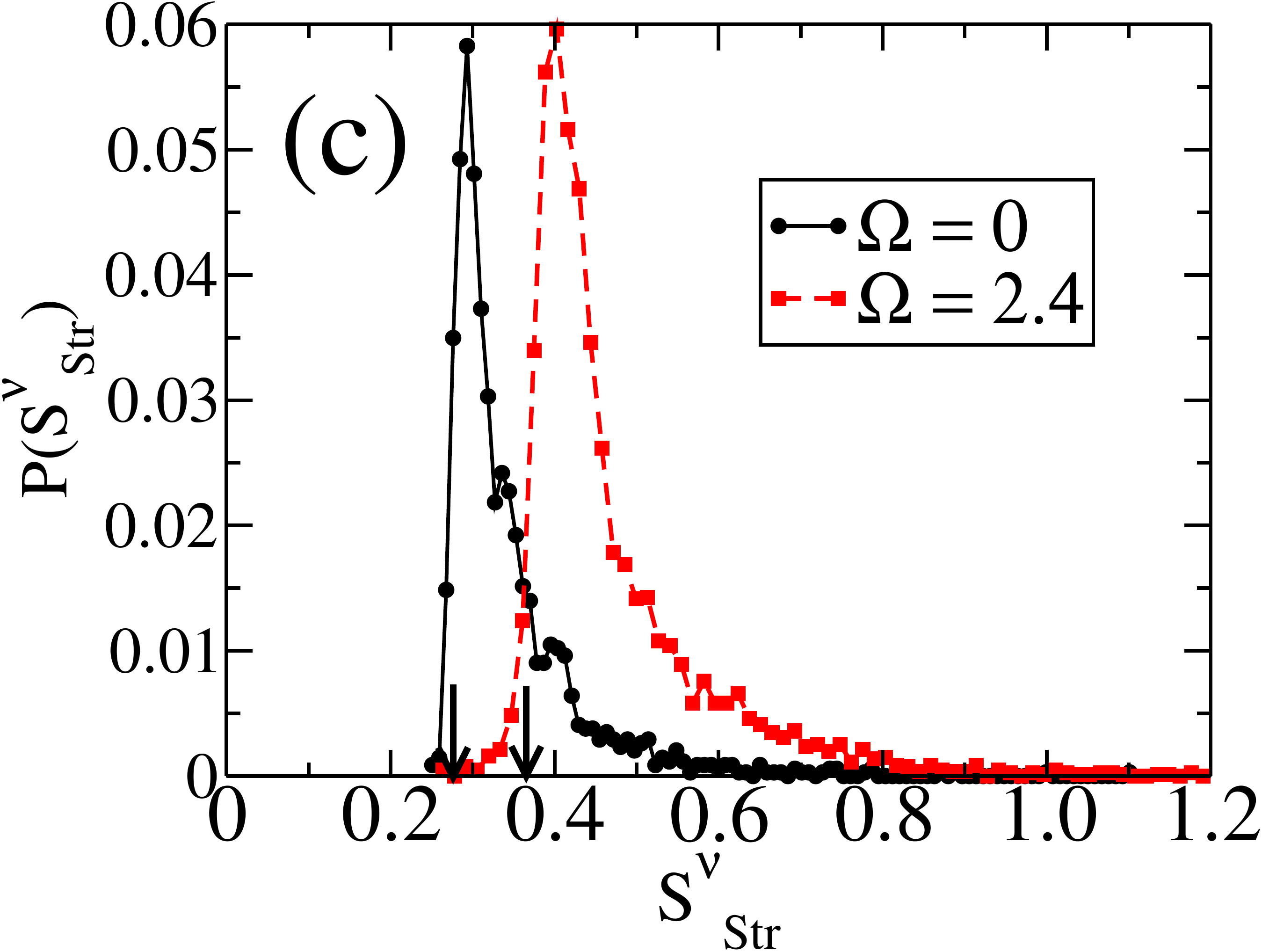}
\includegraphics[width=4.1cm,height=3.3cm]{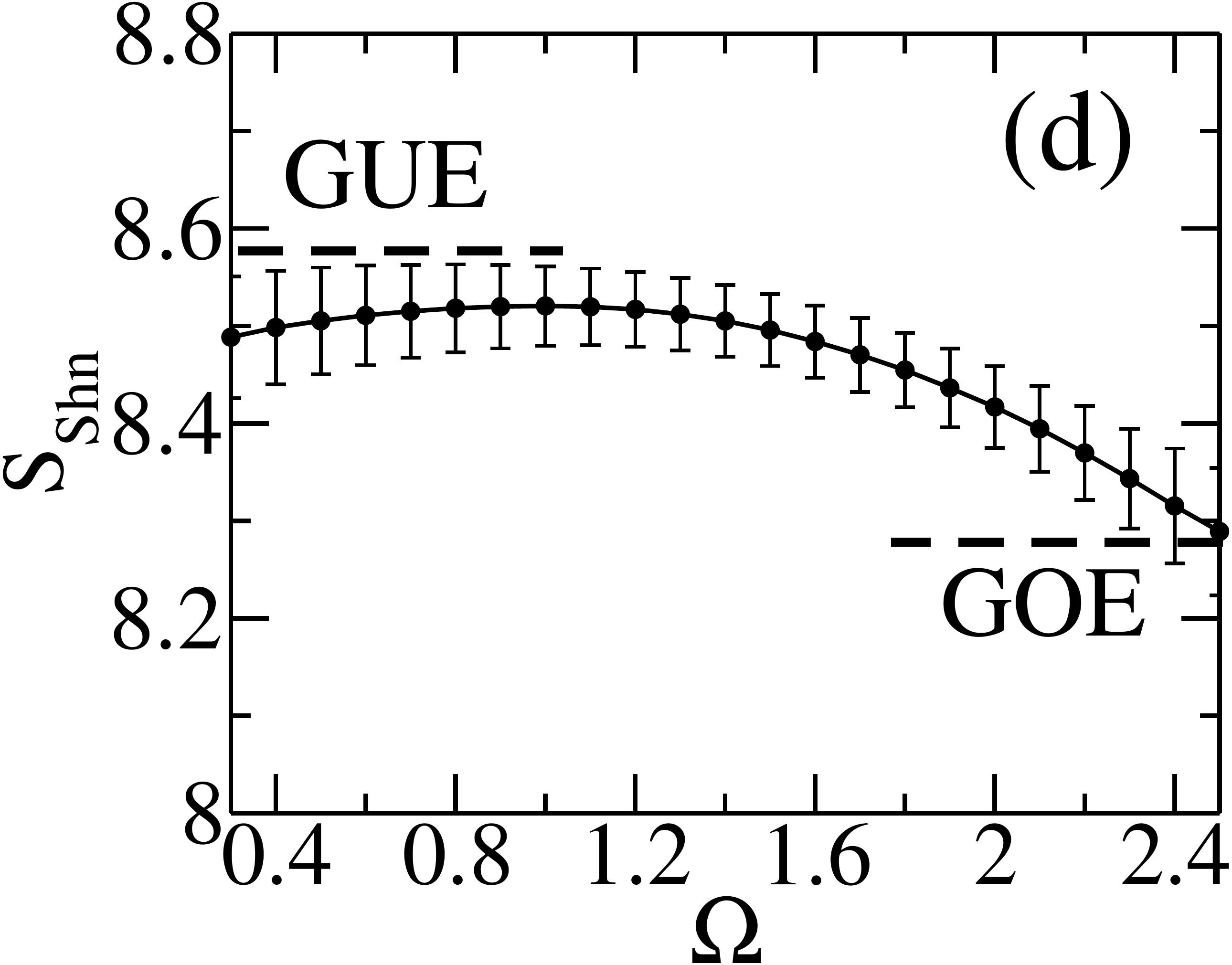}
\caption{(a)-(b) : $\langle r\rangle$ (averaged over 20 disorder
realizations) as a function of $\lambda$ for $\Omega = 0$ and $\Omega
= 2$ respectively. (c) Distribution of $S^{Str}_{\nu}$ of different eigenmodes
$(\nu)$ is shown for $\lambda = 0.8$. Average
$S^{Shn}_{\nu}$ as a function of $\Omega$ is shown in (d) for $\lambda = 0.8$.
We set $V=0.9$ for all the plots.} \label{fig4}
\end{figure}
The phase diagram of the bosons in the large $V/t$ regime as a
function of $\lambda$ can not be completely understood from Fig.\
\ref{fig3}(a) and (b) since $\rho_s=0$ for all $\lambda$ in this
regime. To have an understanding of the nature of the boson phase
with increasing $\lambda$, we study the structure factor $S(k) =
4\sum _{l,l^{\prime}=1}^{N_s}e^{ik(l-l^{\prime})}\langle
\hat{n}_l\hat{n}_{l^{\prime}}\rangle/N_s^2$. We first note that in
the limit $\lambda/t << 1$ the ground state forms a DW leading to
$S(\pi)\simeq 1$ and $S(k) \simeq 0$ for $k \ne \pi$ \cite{SPati}.
This DW state is expected to melt with increasing $\lambda$ leading
to a vanishing of peak of $S(k)$ at $k=\pi$. A plot of $S(\pi)$ in
the $\lambda-V$ plane, shown in Fig.\ \ref{fig3}(c), indicates the
melting with increasing $\lambda$. The dynamical signature of such
melting may be obtained by studying boson dynamics following quench
of $\lambda$ across its melting value \cite{supp1}.

{\it GUE-GOE spectral statistics change}: Finally, we show that the
present model with $V \ne 0$ hosts a change of spectral statistics
from GUE-GOE in the superfluid phase at finite $\lambda$. To this
end, we first note that for $\Omega = 0$, $[\hat{H}, \hat{S}_z] = 0$ and the boson
ground state lies in the $S_z^{\rm total}=N_s/2$ sector. However, for
states within this sector, one does not have TRS since $[\hat{H},
\hat{\mathcal T}] \ne 0$ for a fixed $S_z$ sector. Thus for $\Omega =0$
with a fixed $S_z^{\rm max} \ne 0$ sector, one has $[\hat{H}, \hat{S}_z
\hat{\mathcal T}] \ne 0$. In contrast for $\Omega \neq 0$, it is easy to
see using Eq.\ \ref{ham1}, $[\hat{H}, \hat{S}_z \hat{\mathcal{T}}] = 0$. The latter
symmetry is a consequence of invariance of $\hat{H}$ under TRS followed by
a $\pi$ rotation in spin space about the $z$ axis. The presence of
this additional symmetry leads to GOE to GUE crossover as $\Omega$
is turned on and increased \cite{Haake_p,evec_dist}.

To show this, we first calculate the level spacing ratio \cite{quasirev1,Bogomolny}
$r_{\nu} = {\rm min} (\delta _{\nu+1},\delta _{\nu})/{\rm max}(\delta
_{\nu+1},\delta _{\nu})$, where $\delta _{\nu} = E_{\nu + 1} -
E_{\nu}$, $E_{\nu}$ being the $\nu^{\rm th}$ energy eigenvalue. We
compute the quantity $\langle r \rangle =\sum_{\nu}
r_{\nu}/{\mathcal N}$, where ${\mathcal N}$ is the total number of
levels. For $\Omega = 0$, working with the energy levels in the
maximal $S_z$ sector, we find that $\langle r\rangle $ shows a
crossover from its GUE value of $\approx 0.58$ to that for Poisson
statistics $\approx 0.38$ with increasing $\lambda$ (see Fig.\
\ref{fig4}(a)). In contrast, for large $\Omega = 2$, a similar
analysis shows that $\langle r \rangle$ crosses over from its GOE
value of $\approx 0.527$ to Poisson with increasing $\lambda$ (see
Fig.\ \ref{fig4}(b)).

In finite-sized systems with no strict symmetry breaking, the level
statistics can not be captured for small but finite $\Omega$ values.
We therefore concentrate on the variation of the Shannon and
structural entropy for studying the crossover between GUE-GOE
statistics. The
eigenvector corresponding to the $\nu^{\rm th}$ eigenmode can be
written as $\vert \Phi _{\nu}\rangle = \sum _{\chi} c_{\nu}^{\chi}
\vert \chi \rangle$ where $\vert \chi \rangle$ are the basis states.
The corresponding Shannon entropy is given by $S_{\nu}^{\rm Shn} =
-\sum_{\chi} \vert c_{\nu}^{\chi} \vert ^2 \ln \vert c_{\nu}^{\chi}
\vert ^2$. It is well known that $S^{\rm Shn} = \sum_{\nu}
S_{\nu}^{\rm Shn}$ has the value $S_{GOE}^{\rm Shn} = \Psi(N/2+1) -
\Psi(3/2)$ for GOE and $S_{GUE}^{Shn} = \Psi(N+1) - \Psi(2)$ for GUE
\cite{Haake_b,Izrailev_rmt}. Here $\Psi$ is the Digamma function and
$N$ is the system dimension. The structural entropy for the $\nu$th
eigenmode is defined as follows $S_{\nu}^{Str} = S_{\nu}^{Shn} - \ln
\xi _{\nu}$ where $\xi_{\nu}$ is the IPR corresponding to the
$\nu$th eigenmode. It is known that $S_{\nu}^{Str} \approx 0.37
(0.27)$ for GOE(GUE) \cite{Varga,Izrailev_rmt}. In Fig.\
\ref{fig4}(c), we have plotted the distribution of $S_{\nu}^{Str}$
showing that the peak of the distribution shifts from it's GUE value
($\approx 0.27$) to its GOE value ($\approx 0.37$) as $\Omega$ is
changed from $0.6$ to $2.0$. In Fig.\ \ref{fig4}(d), we plot the
variation of $S_{\nu}^{\rm shn}$ showing a smooth crossover from its
value for GUE to that for GOE with increasing $\Omega$.

{\it Discussion}: Apart from a rich phase diagram, our analysis
shows that in a system of 1D ultracold bosons in an optical lattice
the interplay between SO interaction, Raman coupling and AA
potential leads to novel effects, particularly the splitting in the
spin resolved momentum distribution. Above $\Omega_c$, such a split
happens only when both $\lambda$, $q \ne 0$ and may serve as an
indicator of the localization transition. The experimental
verification of this splitting would involve preparing a system of
bosons with SO coupling \cite{nonab1} in the presence of a 1D
bichromatic lattice to model AA potential \cite{inguscio}; finally
the spin-resolved momentum distribution of these bosons can be
measured by usual Stern-Gerlach technique \cite{exp3}. Our prediction
is that such an experiment would observe a spin-split momentum
distribution near the localization transition which increases with
increasing $\lambda$ or $q$. We note that typically experiments are
done with finite lattice sites $N_{s} \sim 12$ \cite{greiner1}; thus
our numerical results are expected to be of direct relevance for
experimental systems. We have also shown that the spectral
statistics of the present model follows Poissonian distribution for
large $\lambda$ indicating localization and hosts a GUE-GOE
crossover as a function of $\Omega$ in the delocalized regime.
Finally we have identified the existence of localized glassy and DW
phases as a result of the interaction and the AA potential.

{\it Acknowledgement}: BM thanks A. Dutta and S. Mukherjee for
discussion.

\newpage

\begin{widetext}
\appendix

\begin{center}
{\bf SUPPLEMENTAL MATERIAL}
\end{center}

In this supplementary material we provide additional details on the
single-particle spectrum and the phases that we have obtained in the
non-interacting limit of the spin-orbit (SO) coupled Bosons in
presence of Aubry-Andr\'e(AA) potential in the main text. We also
discuss the ground state properties and the localization transition
for the weakly interacting bosons. Finally, we discuss the dynamics
of hard-core bosons (HCB) which may provide a definitive
experimental signature for the identification of the localized
phases in the strongly interacting regime.

\section{Non-interacting limit}
\label{Non_int}

The single particle Hamiltonian of a spin-orbit(SO) coupled bosonic
system in an optical lattice (as given by Eq.\ 1 of the main text
with $\lambda = \mathcal{V} =0$) can be written as a $2 \times 2$
matrix in the momentum representation as
\begin{equation}
H_{SO} = \left( \begin{array}{cc}
-2\cos(k+q) & \Omega \\
\Omega & -2\cos(k-q)
\end{array}\right)
\label{HSO}
\end{equation}
The energy dispersion of $H_{\rm SO}$ is given by Eq.\ 2 of the main
text. From this dispersion, which provides the expression of the
lower branch of the spectrum $E_k^{-}$, we find that there exists a
critical value $\Omega_c$ above which the ground state is doubly
degenerate and the energy minima shifts to finite momenta $k_0 = \pm
\cos ^{-1}[\cos q \sqrt{1 + \Omega ^2/(4\sin ^2 q)}]$ (see Fig.\
\ref{disp}).
\begin{figure}[ht]
\centering
\includegraphics[scale=0.8]{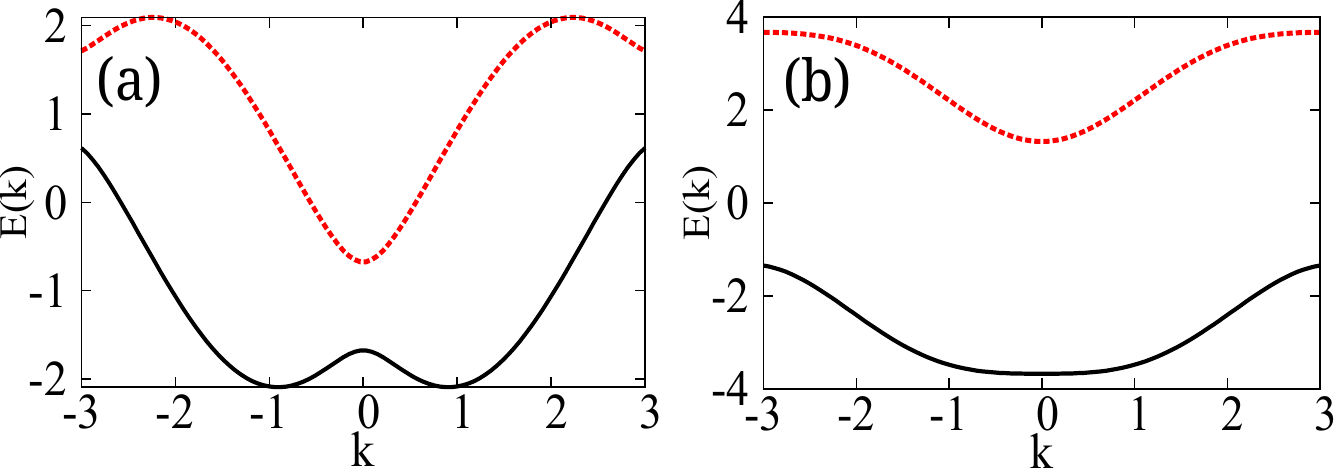}
\caption{Energy dispersion for (a) $\Omega < \Omega _c$ and (b) $\Omega > \Omega _c$}
\label{disp}
\end{figure}
The effective mass (or the band mass) of the bosons is thus given by
$m^* = \partial ^2_k E_k^{-} |_{k=k_0}$. For $\Omega > \Omega _c$,
the expression of $m^{\ast}$ can be written as,
\begin{equation}
m^{\ast}_{\Omega >\Omega_c} \equiv m^*_{>} = \left(\frac{\partial ^2
E_k^{-}}{\partial k^2}\right)_{k=0} = (1-\Omega _c/\Omega)\cos q.
\end{equation}
We note that the effective masses $m^*_{>}(m^*_{<})$ in the two
regimes $\Omega > \Omega_c(<\Omega_c)$ both vanish at
$\Omega=\Omega_c$. Furthermore, in the absence of disorder the
superfluid fraction (SFF) turns out to be the boson effective mass.
Thus $m^{\ast}$ captures the behavior of the SFF obtained
numerically as a function of $\Omega$ (see Fig.\ \ref{SFF}(a)); this
situation is similar to that obtained in the continuum limit
\cite{stringari}.

In presence of the AA potential we numerically diagonalize the
single particle Hamiltonian to obtain the ground state and the full
excitation spectrum. The localization transition of the ground state
is characterized by the vanishing of superfluid fraction (SFF) which
can be obtained using Eq.\ 4 of the main text. Alternatively, one
can also adopt the perturbative approach to calculate the SFF. To
this end, we note that in the presence of a small phase twist
$\theta$ across the boundary, the original Hamiltonian becomes,
\begin{equation}
H_{\theta} = -t\sum_{l,\sigma}
\left(\hat{b}_{l,\sigma}^{\dagger}e^{iq\hat{\sigma}_z}\hat{b}_{l+1,\sigma}e^{-i
\theta /N_s}+h.c.\right)+\Omega
\sum_{l,\sigma}\hat{b}_{l,\sigma}^{\dagger}\hat{\sigma}_x\hat{b}_{l,\sigma}
+ \lambda \sum_{l,\sigma}\cos(2\pi \beta
l)\hat{b}_{l,\sigma}^{\dagger}\hat{b}_{l,\sigma}. \label{ptwist}
\end{equation}
An expansion of the $H_{\theta}$ (Eq.\ \ref{ptwist}) to ${\rm
O}(\theta^2)$ yields,
\begin{equation}
H_{\theta} = H_{0} + \frac{\theta}{N_s}\hat{J} - \frac{\theta ^2}{2N_s^2}\hat{T}
\end{equation}
where, $H_{0}$ is the unperturbed Hamiltonian, $\hat{T} = -t\sum
_{l,\sigma} (\hat{b}_{l+1,\sigma}^ {\dagger} e^{iq\hat{\sigma}_z}
\hat{b}_{l+1,\sigma} + h.c.)$ is the kinetic energy operator and
$\hat{J} = it\sum _{l,\sigma} (\hat{b}_{l+1,\sigma}^
{\dagger}e^{iq\hat{\sigma}_z}\hat{b}_{l,\sigma} - h.c.)$ is the
current operator. So, to $O(\theta^2)$ the superfluid fraction is
given by,
\begin{equation}
f_s = -\frac{1}{2t}\langle \psi_0 \vert \hat{T} \vert \psi_0 \rangle
- \frac{1}{t}\sum _{\nu \ne 0} \frac{|\langle \psi_{\nu} \vert
\hat{J} \vert \psi_0 \rangle |^2}{E^{\nu} - E^{0}}
\end{equation}
where $0$ and $\nu$ stands for the lowest and $\nu$th eigenmode
respectively. In Fig.\ \ref{SFF}(b) we have plotted $f_S$ using the
above prescription; we note that the SFF vanishes at $\lambda_c < 2$
at which the IPR starts rising indicating the localization
transition.

The localization transition can also be qualitatively understood
from the vanishing of the energy gap $\Delta E$ at the critical
disorder strength $\lambda _c$. The energy gap ($\Delta E$) between
the ground state and the 1st excited state is expected to vanish at
the localization transition point. Using this fact, one can obtain a
qualitative understanding of the phase diagram for both small and
large $\Omega$. We note that for small $\delta \Omega =
\Omega-\Omega_c >0$ at $\lambda=0$, the ground state is at $k=0$ and
has an energy $E(k=0) = -(\Omega + 2 \cos q)$. Now let us turn on
$\lambda$ which leads to a perturbation term, which can be written
in momentum space as
\begin{eqnarray}
H_1 = \frac{\lambda}{2} \sum_k {\hat b}_{k \sigma}^{\dagger} ({\hat
b}_{k + 2\pi \beta \, \sigma} + {\hat b}_{k - 2\pi \beta \,
\sigma}).
\end{eqnarray}
Such a perturbation term leads to a hybridization of the ground
state at $k=0$ with the one at $k=\beta$ which has energy
$E(k=\beta) = -2 \cos \beta \cos q - |2 \sin \beta \sin q| + {\rm
O}(\Omega^2)$. Thus the simplest qualitative estimate of the
transition line for small $\delta \Omega$ occurs when $\lambda
\simeq E(k =\beta)-E(k=0)$ leading to
\begin{eqnarray}
\lambda= \delta \Omega (1 - \tan q/\sqrt{\sin^2 \beta \sin^2 q}) + 2
\cos q(1-\cos \beta) + |2 \sin q \tan q| -|\sin(q)| \sqrt{\sin^2
\beta + \tan^2 q}.
\end{eqnarray}
We note that this reproduces the linear behavior of the phase
boundary for small $\delta \Omega$. A similar analysis can also be
carried out at $\Omega \gg 1$. Here the ground state is again at
$k=0$ for $q < \pi/2$. An exactly similar analysis as the one
charted out above shows that for this case $E[k=\beta]-E[k=0] = 2
\cos q(1-\cos \beta) + {\rm O}(1/\Omega)$ which leads to
\begin{eqnarray}
\lambda \simeq 2 \cos q(1-\cos \beta)
\end{eqnarray}
Thus the phase boundary becomes a horizontal line in the
$\lambda-\Omega$ plane, as also seen in exact numerics. In Fig.\
\ref{SFF}(c) we have shown $\Delta E$ as a function of $\lambda$ for
two different system sizes corroborating these qualitative features
and justifying the assumption of vanishing of $\Delta E$ at the
transition point.
\begin{figure}[ht]
\centering
\includegraphics[scale=0.2]{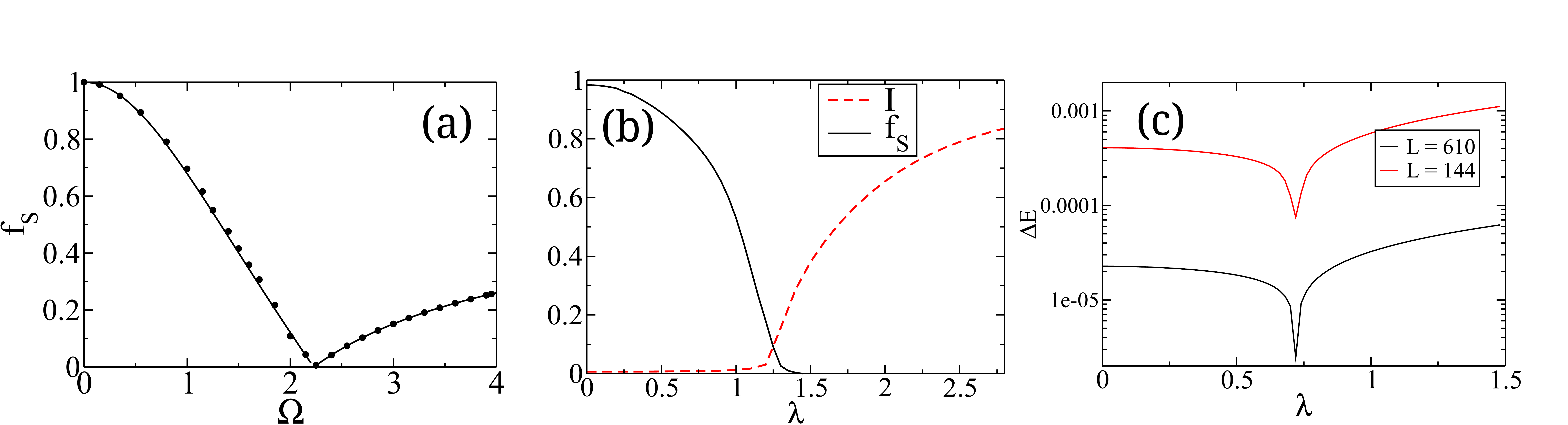}
\caption{(a) Superfluid fraction as a function of $\Omega$ for
$\lambda = 0$. Solid line represents the SFF obtained analytically
from the effective mass calculation. (b) SFF and IPR is plotted as a
function of $\lambda$ for $\Omega = 0.2$. (c) Energy gap between
ground state and 1st excited state as a function of $\lambda$ for
$\Omega = 3.5$.} \label{SFF}
\end{figure}

\section{Localization of weakly interacting bosons} \label{weak_int}

In the weakly interacting limit, {\it i.e.}, for $U/t<<1$ and $V=0$,
we replace the quantum field operator $\hat{b}_{l,\sigma}$ by the
classical field operator $\psi _{l,\sigma}$ assuming the existence
of a 1D quasi-condensate \cite{shlyapnikov1d}. By minimizing the
energy functional calculated thereby, we obtain the discrete
non-linear Schr\"odinger (DNLS) equation for the condensate wave
function $\psi_{l,\sigma}$ given by,
\begin{eqnarray}
&&-(\psi _{l+1,\uparrow}e^{iq} + \psi _{l-1,\uparrow}e^{-iq})
+ \lambda \cos (2\pi \beta l) \psi _{l,\uparrow} +
\Omega \psi _{l,\downarrow} \nonumber + U(|\psi _{l,\uparrow}|^2
+ |\psi _{l,\downarrow}|^2)\psi _{l,\uparrow} = \mu \psi _{l,\uparrow} \\
&&-(\psi _{l+1,\downarrow}e^{-iq} + \psi _{l-1,\downarrow}e^{iq})
+ \lambda \cos (2\pi \beta l) \psi _{l,\downarrow} +
\Omega \psi _{l,\uparrow} \nonumber + U(|\psi _{l,\uparrow}|^2
+ |\psi _{l,\downarrow}|^2)\psi
_{l,\downarrow} = \mu \psi _{l,\downarrow}
\end{eqnarray}
where $\mu$ is the chemical potential. We then obtain the ground
state wavefunction $\psi_{l \sigma}$ numerically and use it to
compute all relevant quantities such as IPR and $f_s$. In what
follows we have shown the results of such numerical study which are
shown in Fig.\ \ref{PD_int}.

\begin{figure}[ht]
\centering
\includegraphics[scale=0.2]{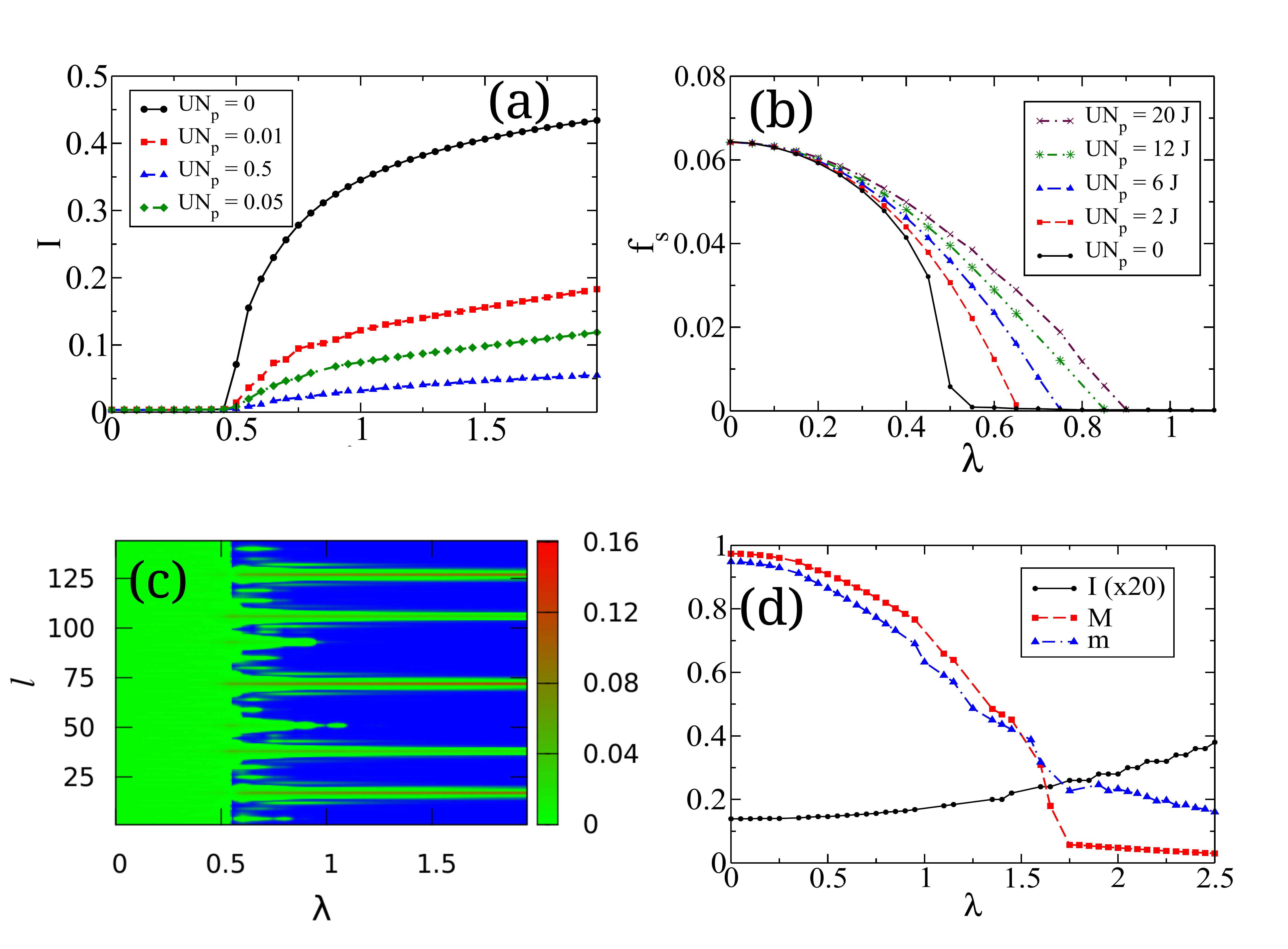}
\caption{The IPR and SFF has been shown as a function of $\lambda$
for $\Omega = 2.5$ and for different interaction strength $UN_p/t$
in (a) and (b) respectively. The spatial distribution of the ground
state density has been shown for $UN_p = 0.5$, $\Omega = 2.5$ in
(c). IPR, the order parameter $m$ and the total magnetization $M$ as
a function of $\lambda$ for $UN_p = 20$ and $\Omega = 0.3$ are shown in (d).
We set $N_p=200$ and $N_s=144$ for all the plots.}
\label{PD_int}
\end{figure}

In Fig.\ \ref{PD_int}(a) we plot the ground state IPR  as a function
of $\lambda$ for different interaction strength $UN_p/t$. We see
that on increasing $\lambda$ beyond the localization transition, the
growth of IPR decreases. This is due to the fact that the ground
state wavefunction becomes multi-site localized due to weak
repulsive interaction (see Fig.\ \ref{PD_int}(c)). We further
calculate the superfluid fraction which vanishes in the localized
phase as depicted in Fig.\ \ref{PD_int}(b).

\begin{figure}[ht]
\centering
\includegraphics[scale=0.3]{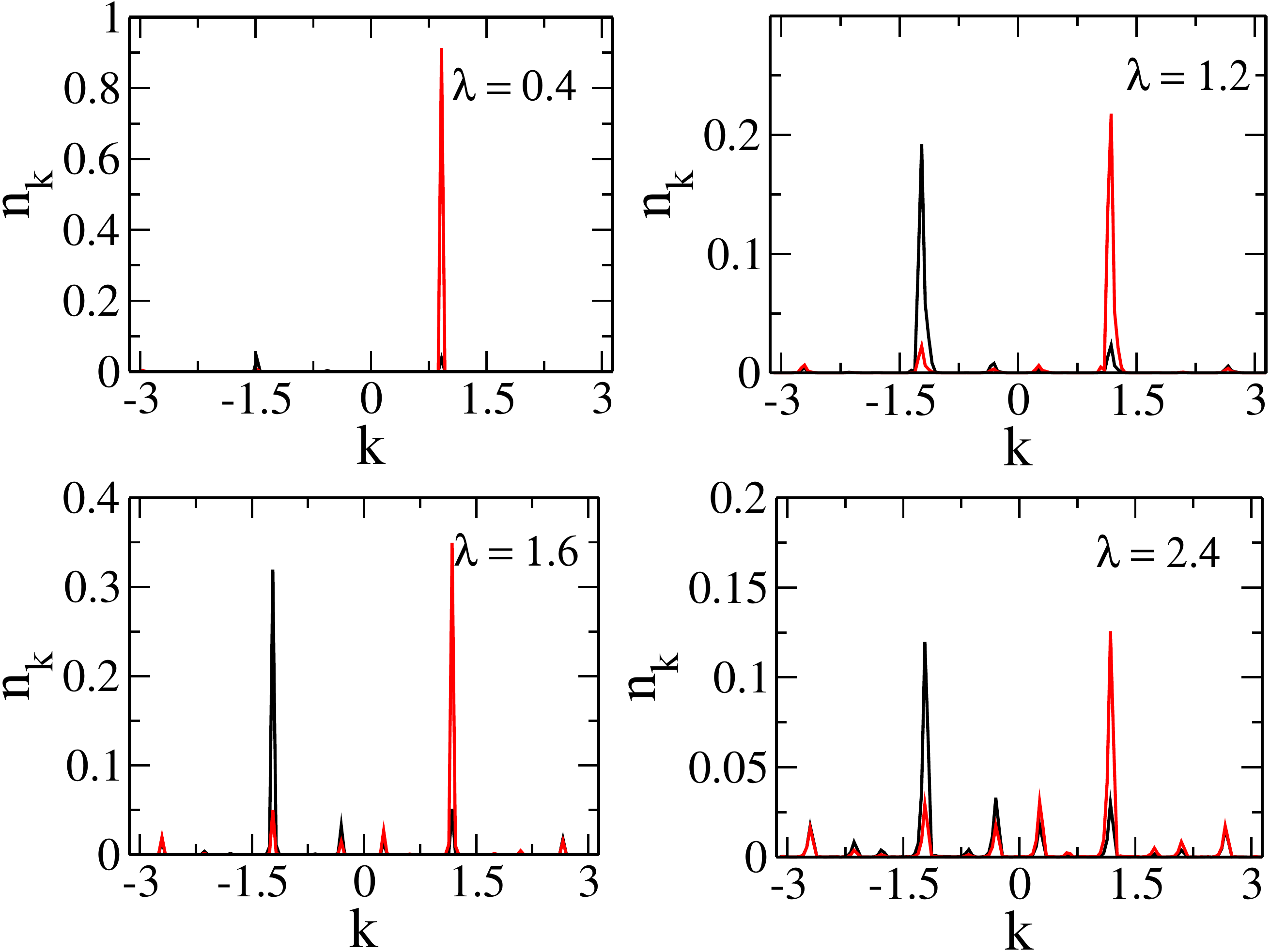}
\caption{Momentum distribution with increasing disorder strength
$\lambda$ for $\Omega = 0.5$ and $UN_p = 20$.} \label{k_dist_int}
\end{figure}

To gain a better understanding of the localization transition, we
further study the spin resolved momentum distribution of bosons in
the regime $\Omega < \Omega _c$. In contrast to the non-interacting
case, the superfluid with finite $U$ the chooses one of the two
symmetry broken states with spins polarized along the
z-axis \cite{stringari1}. As a result the momentum distribution
corresponding to the spin polarization of the ground state becomes
highly peaked at the nonvanishing momentum of the ground state as
depicted in Fig.\ \ref{k_dist_int}(a). With increasing disorder
strength, other momentum modes get gradually occupied and
spin-momentum distributions are peaked at equal and opposite
momentum with a net spin polarization indicating symmetry breaking
(see Fig.\ \ref{k_dist_int}). Finally in the localized phase, the
momentum distributions become symmetric and peaked around finite
momentum with $n_{k,\uparrow} = n_{-k,\downarrow}$. To verify this
we plot the order parameter $m = \sum_k (n_{k,\uparrow} - n
_{-k,\downarrow})^2$ and the total magnetization $M =
\sum_k(n_{k,\uparrow} - n _{k,\downarrow})$ which decreases with
increasing $\lambda$ and finally vanishes in the localized phase
(see Fig.\ \ref{PD_int}(d)).

Next we investigate the momentum distribution in the regime $\Omega
> \Omega_c$; similar to the non-interacting case, we see that in the
delocalized regime the momentum distributions for both up and down
spins are peaked at zero momentum, whereas, in the localized phase
they are peaked at finite momentum and other momentum modes get
gradually occupied (see Fig.\ \ref{phase_diff_int}(a,b)).

\begin{figure}[ht]
\centering
\includegraphics[scale=0.18]{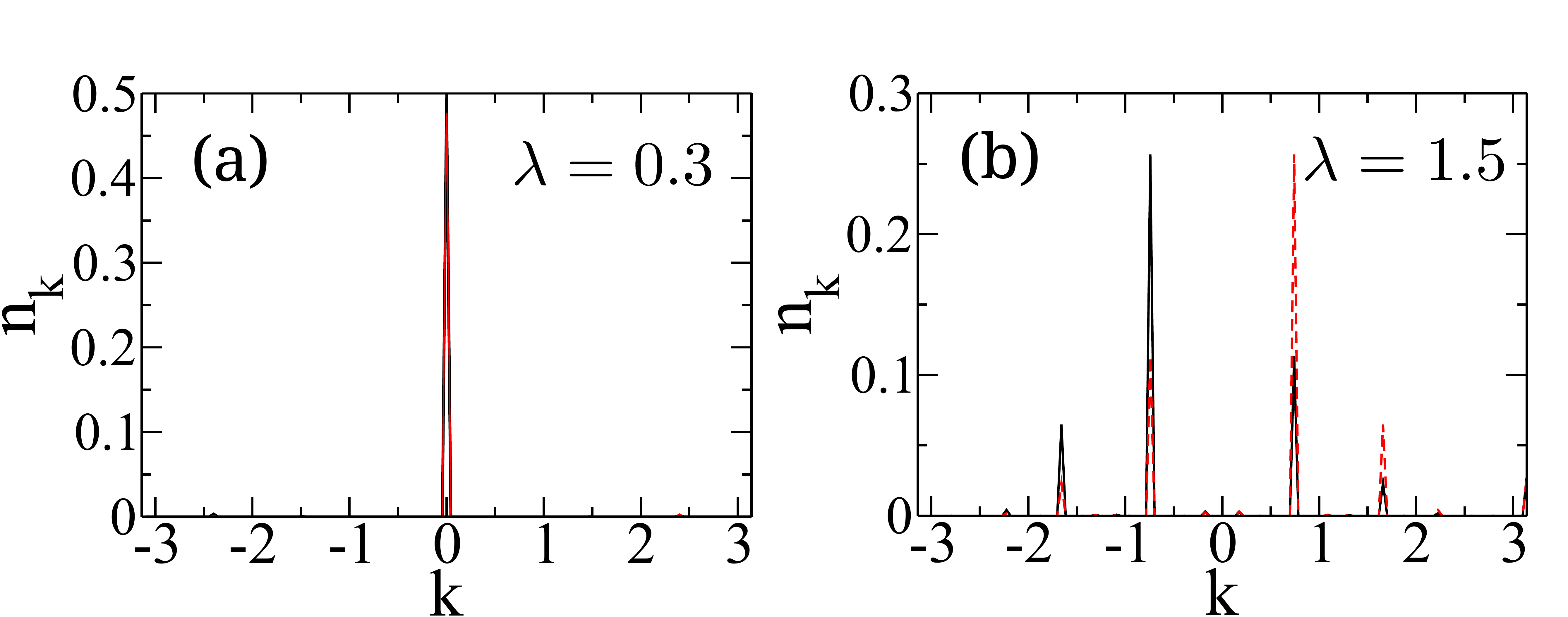}
\caption{(a)-(b) Momentum distribution for up(down) spin is shown in
solid(dashed) line with increasing disorder strength $\lambda$.
Other parameters are $\Omega = 2.5$ and $UN_p = 10$.}
\label{phase_diff_int}
\end{figure}

\section{Peak splitting and spin dephasing near the localization
transition}

This effect of spin-splitted momentum distribution of the localized
wavefunction in the regime $\Omega >\Omega_c$ arises due to the
interplay between the SO interaction and Raman coupling. Here we
provide a simple variational calculation to understand this effect.
First we consider the variational wavefunction given by
\begin{equation}
\psi_l = \mathcal{N}e^{-|l|/\xi}\left(\begin{array}{c} e^{ikl}\\
-e^{-ikl} \\ \end{array}\right), \quad \mathcal{N} =
\sqrt{\frac{\tanh(1/\xi)}{2}}
\end{equation}
where $l$ is the site index and $\xi$ represents localization length
which is assumed to subsume the effect of AA potential and the
interaction. The spinor part is chosen in such a way that up(down)
spin momentum distribution is peaked at $\pm k$ and for $k=0$ it
reduces to the usual form of the ground state for $\Omega >
\Omega_c$. For the aforementioned wavefunction, the parameter $k$ is
treated as the variational parameter and we investigate its
dependence on $\xi$ and $\Omega$. Considering the single particle
Hamiltonian of a spin-orbit(SO) coupled bosonic system in an optical
lattice (Eq.\ 1 of the main text with $\lambda = \mathcal{V} =0$),
the energy can be written as,
\begin{equation}
E(k) = - \left[\frac{\cos(k-q)}{\cosh(1/\xi)} + \Omega
\tanh(1/\xi)\frac{\sinh(2/\xi)}{\cosh(2/\xi) - \cos 2k}\right].
\end{equation}
From the structure of $E(k)$ we note that the $E(k) \to - (\cos(k-q)
+\Omega \delta_{k0}) +{\rm O}(1/\xi)$ in the delocalized limit where
$\xi \to \infty$. This implies that this functional reproduces the
correct $k=0$ ground state for $\Omega > \Omega_c$ in the absence of
the AA potential. Thus in this case the momentum distribution of
both the spin-up and spin-down components are peaked at $k=0$. In
the strongly localized phase, were $\xi \ll 1$, the second term
dominates and in the limit of single site localization $k$ looses
its meaning. However, in between these two limits for finite $\xi$,
the ground state minima shifts to finite $k$ provided $q \ne 0$.
This is seen by minimizing $E(k)$ to obtain $k_{min}$ and by
plotting its variation as a function of $\xi^{-1}$ as shown in Fig.\
\ref{split_peak}. As seen from Fig.\ \ref{split_peak}, it is evident
that $k_{min}$ decreases with increasing $\xi$ and finally it
vanishes in the delocalized regime i.e. $\xi^{-1} \rightarrow 0$. We
further notice for a fixed $\xi$ the spin splitting (characterized
by $k_{min}$) decreases with decreasing strength of SO interaction
($q$) and eventually vanish for $q = 0$. This simple variational
calculation elucidates how the combined effect of localization and
SO interaction gives rise to the spin splitted momentum distribution
in the regime $\Omega > \Omega_c$.

\begin{figure}[ht]
\begin{center}
\includegraphics[scale=0.2]{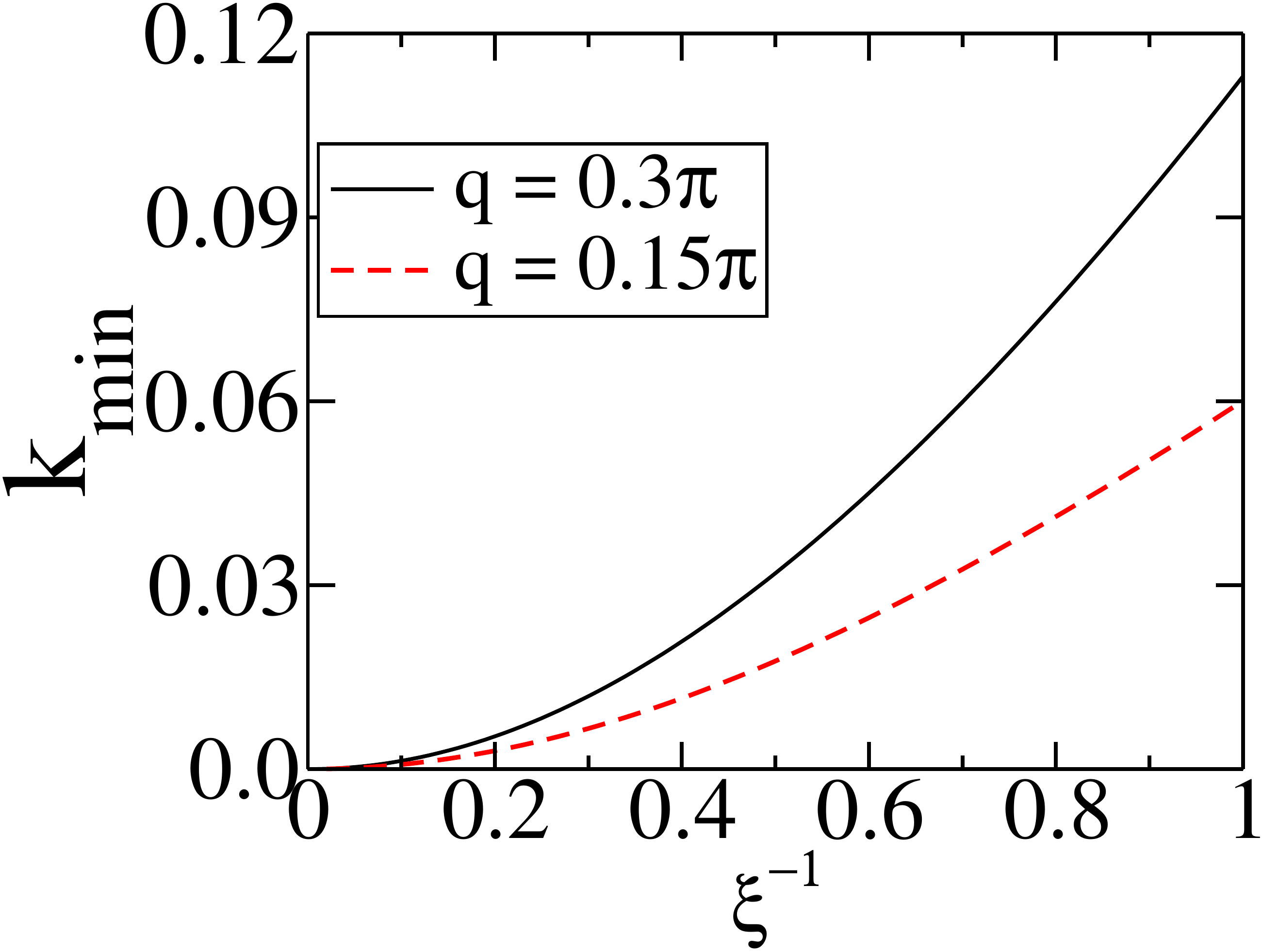}
\end{center}
\caption{$k_{min}$ as a function of $\xi ^{-1}$ is plotted for $\Omega = 3$.}
\label{split_peak}
\end{figure}

{\it Spin-phase diffusion}: The effect of spin-splitting in momentum
distribution near the localization transition is also accompanied
with the phase fluctuation of the wavefunction. In general, the
wavefunction can be written as,
\begin{equation}
\psi^{l} = \sqrt{n_{0}^{l}}\left(\begin{array}{c} \cos \theta^l e^{i\phi _{\uparrow}^{l}}\\ \sin \theta^l e^{i\phi _{\downarrow}^{l}} \\ \end{array}\right)
\end{equation}
where $\phi _l = \phi _{\uparrow}^{l} - \phi _{\downarrow}^{l}$ is
the phase angle of the spinor at site $l$. For $\Omega > \Omega_c$,
we find that $\cos \theta = \sin \theta \approx 1/\sqrt{2}$ and
$\phi \approx \pi$ in the delocalized phase, whereas, near
localization transition due to increasing phase fluctuations, the
phase angle fluctuates significantly from $\pi$ at different sites.
We quantify the phase fluctuation by calculating $|\langle e^{i\phi}
\rangle |$,  where the average is taken over all the lattice sites.
In Fig.\ \ref{spin_diff_hc} we have shown the behavior of $|\langle
e^{i\phi} \rangle |$ as a function of the disorder potential
strength $\lambda$ which shows that near the localization transition
it decreases from $1$ with increasing strength of the disorder
$\lambda$. In case of hard core bosons we consider the eigenvector
corresponding to the largest eigenvalue of the density matrix
defined in the main text and calculate the similar quantity.

\begin{figure}[ht]
\begin{center}
\includegraphics[scale=0.2]{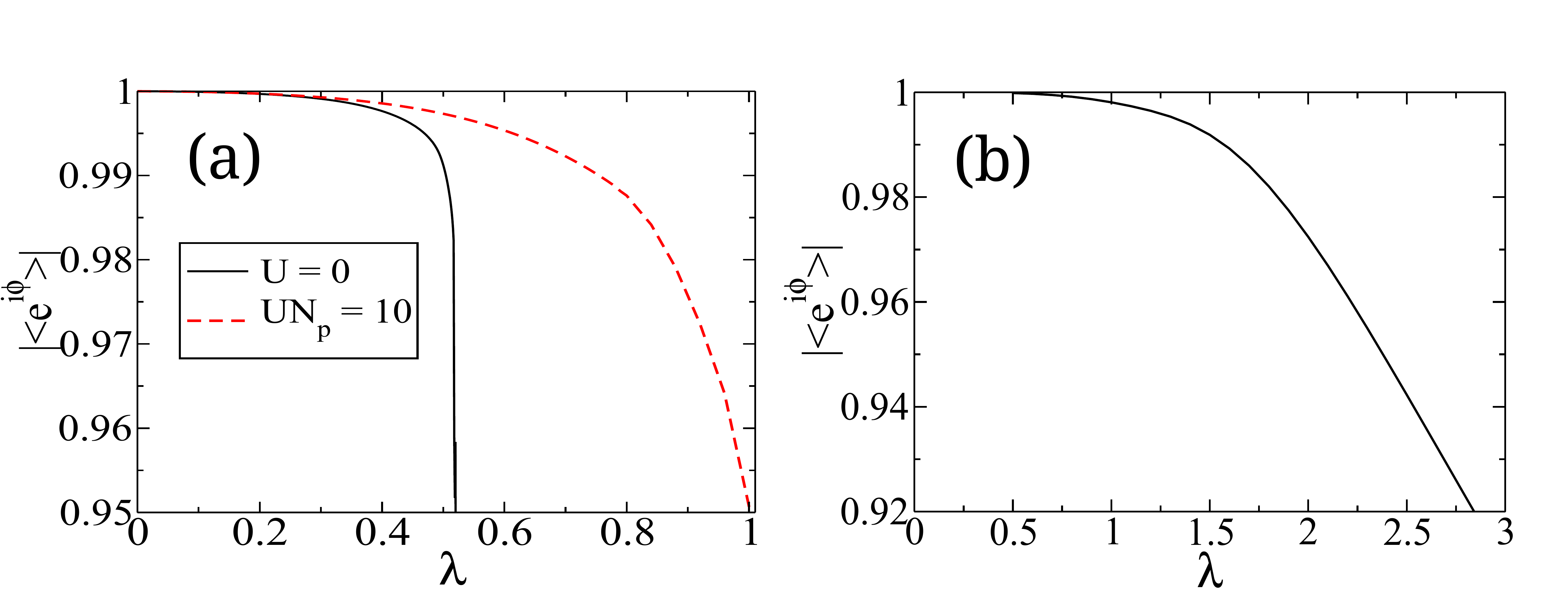}
\end{center}
\caption{$|\langle e^{i\phi} \rangle |$ as a function
of $\lambda$ has been plotted for (a) non-interacting and weakly interacting bosons,
(b) for hard core bosons. Other parameters taken are $\Omega = 2.5$ and $q = 0.3\pi$.} \label{spin_diff_hc}
\end{figure}

\section{Non-equilibrium dynamics in the strongly interacting regime}
\label{dynamics}



To elucidate the localization transition of the HCB,
we now look at into the non-equilibrium dynamics of the bosons.
We start from the density wave state at $\lambda=0$
denoted by $|\psi(0)\rangle$. Next we quench
$\lambda$ to a finite value $\lambda_f$ so that the system
Hamiltonian after the quench is given by $H[\lambda_f]$. Let us
denote the eigenfunctions and eigenvalues of $H[\lambda_f]$ as
$|m\rangle$ and $\epsilon_m$ respectively. The time evolved wavefunction
$|\psi(t)\rangle$ at any instant of time $t$ after the quench can be
obtained by solving the Schrodinger equation $i \hbar
|\psi(t)\rangle = H[\lambda_f] |\psi(t)\rangle$ and is given by
\begin{eqnarray}
|\psi(t)\rangle &=& \sum_{m} c_m e^{-i \epsilon_m t/\hbar}
|m\rangle, \quad c_m = \langle m|\psi(0)\rangle \label{qdyn1}
\end{eqnarray}
The expectation value of any operator $O(t)$ can be obtained from
$|\psi(t)\rangle$ as
\begin{eqnarray}
\langle \psi(t)|O|\psi(t) \rangle = \sum_{m,n} c_m^{\ast} c_n
e^{i(\epsilon_m -\epsilon_n)t/\hbar} \langle m |O| n\rangle
\label{qdyn2}
\end{eqnarray}
Using Eq.\ \ref{qdyn2}, we calculate the time evolution of the
imbalance factor which is defined as,
\begin{equation}
\mathcal{I} = \frac{N_o-N_e}{N_{tot}}
\end{equation}
where, $N_{o[e]} = \langle \psi(t)|\sum_{ r \in {\rm odd[even]
sites}} {\hat b}_{i}^{\dagger} {\hat b}_i|\psi(t)\rangle$ and
$N_{tot}= N_o + N_e$. Note that at $t=0$, we have density wave state
with ${\mathcal I} =1$ and it approaches zero for a delocalized
state. In Fig.\ \ref{Imbalance}(a) we have shown the time evolution
of $\mathcal{I}(t)$ for the up spin species (the same feature can be
observed for the down spin species as well) for different $\lambda$
values. We note that for small $\lambda$ which corresponds to the
delocalized regime, $\mathcal{I}$ vanishes to zero with time showing
the ergodic dynamics in that regime, whereas for larger $\lambda$
value which corresponds to the localized regime, $\mathcal{I}$
doesn't vanish and saturates to some positive value which indicates
the non-ergodic regime and the density wave ordering is retained in
the course of time evolution. In Fig.\ \ref{Imbalance}(b) the final
density distribution after the time evolution has been shown for
different values of $\lambda$.

We repeated the same numerical experiment starting from a different
initial state where the atoms are loaded on one half of the lattice
and study the imbalance factor $\mathcal{I'} = (N_l-N_r)/N_{tot}$ as
a function of time $N_l$ and $N_r$ being the total number density of
bosons at the left and the right halves of the lattice respectively.
In Fig.\ \ref{Imbalance}(c,d) we have plotted the time evolution of
the imbalance factor and the final density distribution of the up
spin species for different values of the disorder strength.

\begin{figure}[ht]
\centering
\includegraphics[scale=0.2]{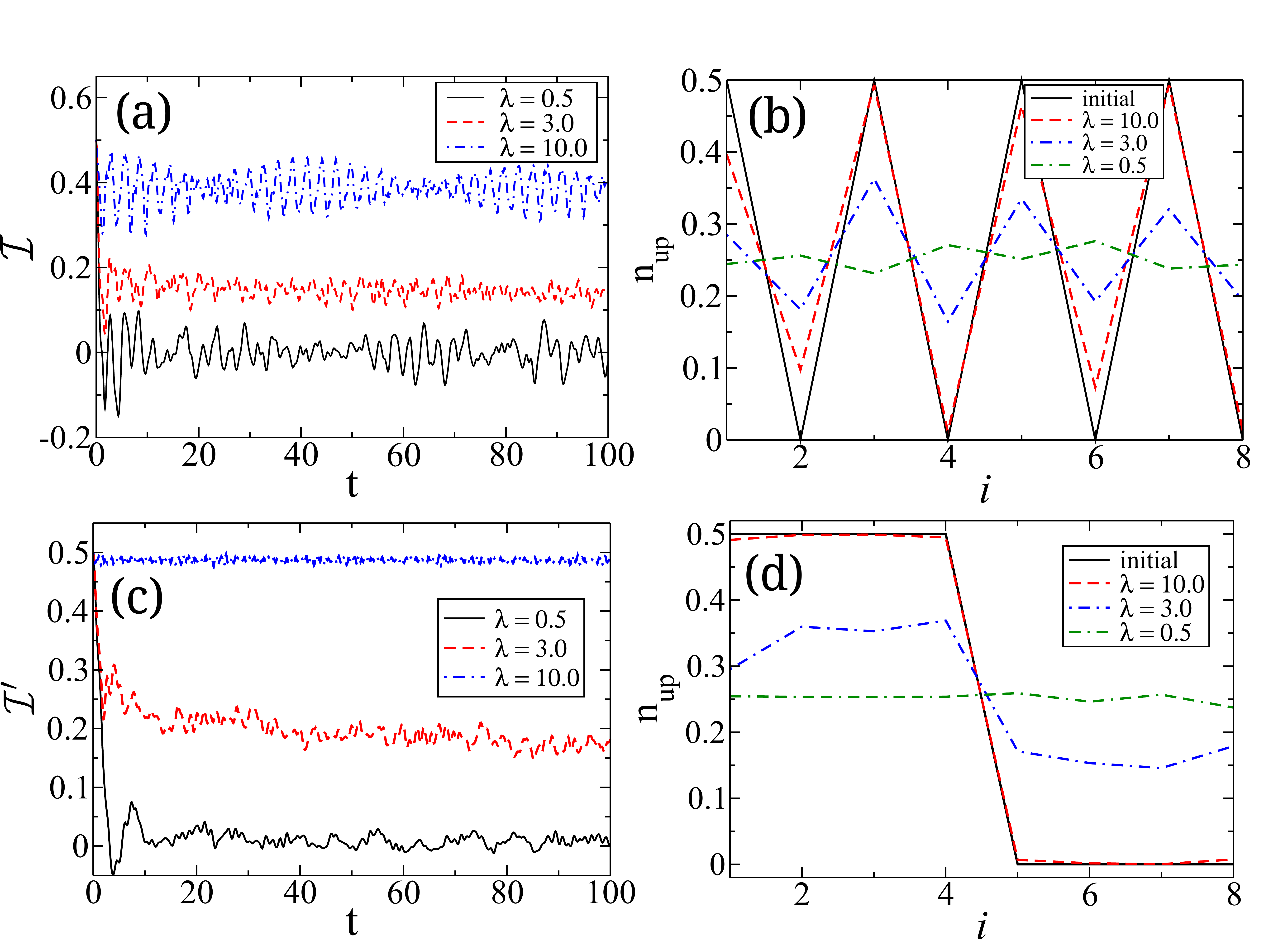}
\caption{The time evolution of the imbalance factor for up spin
species has been shown starting from two initial states (a) density
wave and (c) bosons loaded in left half of the lattice for different
values of the disorder strength $\lambda$. The final up spin density
distribution at the end of the time evolution for the same
$\lambda$'s are shown for the two types of initial states in (b) and
(d) respectively. The other parameters are $\Omega = 2$, $q =
0.3\pi$ and $V = 0$.} \label{Imbalance}
\end{figure}

\end{widetext}


\begin{thebibliography}{99}

\bibitem{mbl1} 
R. Nandkishore and D. A. Huse, Annu. Rev. Condens. Matter Phys. {\bf 6}, 15 (2015); E. Altman and R. Vosk, Annu. Rev. Condens. Matter Phys. 6, 383 (2015)

\bibitem{mbl2} A. Pal and D. A. Huse, \prb {\bf 82}, 174411 (2010); I. L. Aleiner, B. L. Altshuler and G. V. Shlyapnikov, {\it Nature Physics} {\bf 6}, 900-904 (2010).

\bibitem{boserev1} M. Greiner, O. Mandel, T. Esslinger, T. W. Hänsch and I. Bloch, {\it Nature} {\bf 415}, 39-44 (2002); I. Bloch, J. Dalibard and W. Zwerger, \rmp {\bf 80}, 885 (2008).

\bibitem{boserev2} M. Schreiber {\it et al}, Science {\bf 349}, 842 (2015);
P. Bordia, H. P. L\"uschen, S. S. Hodgman, M. Schreiber, I. Bloch
and U. Schneider, \prl {\bf 116}, 140401 (2016); S. S. Kondov, W. R.
McGehee, W. Xu and B. DeMarco, \prl {\bf 114}, 083002 (2015).

\bibitem{quasiref1} A. I. Goldman and R. F. Kelton, \rmp {\bf 65}, 213 (1993).

\bibitem{quasiref2} R. Lifshitz, \rmp {\bf 69}, 1181 (1997).

\bibitem{quasiref3} M. Quilichini, \rmp. {\bf 69}, 277 (1997).

\bibitem{aapaper} S. Aubry and G. Andr\'e, Ann. Israel. Phys. Soc. {\bf 3}, 133 (1980)

\bibitem{aapaper1} C. Aulbach, A. Wobst, G. L. Ingold, P. H\"anggi and I. Varga, New J Phys, {\bf 6}, 70 (2004); M. Modugno, New J. Phys. {\bf 11}, 033023 (2009).

\bibitem{quasirev1} S. Iyer, V. Oganesyan, G. Refael and D. A. Huse, \prb {\bf 87}, 134202 (2013).

\bibitem{mblaa2} X. Li, S. Ganeshan, J. H. Pixley, and S. D. Sarma, \prl {\bf 115}, 186601 (2015).

\bibitem{loclight1} Y. Lahini, R. Pugatch, F. Pozzi, M. Sorel, R. Morandotti, N. Davidson, and Y. Silberberg, \prl {\bf 103}, 013901 (2009).

\bibitem{inguscio} G. Roati {\it et al}, {\it Nature} {\bf 453}, 895 (2008).

\bibitem{kushref1}  K. Singh, K. Saha, S. A. Parameswaran, and D. M. Weld,
\pra {\bf 92}, 063426 (2015).

\bibitem{localization} G. Modugno, Rep. Prog. Phys. {\bf 73}, 102401 (2010); V. P. Michal,
B. L. Altshuler, and G. V. Shlyapnikov, \prl {\bf 113}, 045304 (2014);
S. Ray, M. Pandey, A. Ghosh and S. Sinha, New J. Phys. {\bf 18}, 013013 (2016).

\bibitem{inguscio1}
L. Fallani, J. E. Lye, V. Guarrera, C. Fort and M. Inguscio, \prl {\bf 98}, 130404 (2007);
\prl 113, 095301 (2014); Chiara D'Errico {\it et al}, \prl {\bf 113}, 095301 (2014).

\bibitem{demarco}
M. White, M. Pasienski, D. McKay, S. Q. Zhou, D. Ceperley, and B. DeMarco, \prl {\bf 102}, 055301 (2009);
C. Meldgin,   U. Ray, P. Russ, D. Chen, D. M. Ceperley and B. DeMarco, {\it Nat. Phys.} {\bf 12}, 646 (2016).

\bibitem{bhref1}
M. P. A. Fisher, \prb {\bf 40}, 546 (1989); G. Roux, T. Barthel, I.
P. McCulloch, C. Kollath, U. Schollwck, and T. Giamarchi, \pra {\bf
78}, 023628 (2008);  G. Roux, A. Minguzzi and T. Roscilde, New J.
Phys. {\bf 15}, 055003 (2013).

\bibitem{Roth} R. Roth and K. Burnett, \pra {\bf 68}, 023604 (2003).

\bibitem{ab1} Y.-J. Lin, R. L. Compton, K. Jimenez-Garcia, J. V. Porto, and I. B.
Spielman, Nature (London) {\bf 462}, 628 (2011).

\bibitem{nonab1} J. Dalibard, F. Gerbier, G. Juzeli\=unas and P. \"Ohberg, \rmp {\bf 83}, 1523 (2011).

\bibitem{abph1} S. Sinha and K. Sengupta, Europhys. Lett. {\bf 93}, 30005 (2011);
S. Powel, R. Barnett, R. Sensarma, and S. D. Sarma, \prl {\bf 104}, 255303 (2010);
K. Saha, K. Sengupta, and K. Ray, \prb {\bf 82}, 205126 (2010).

\bibitem{abph2} D. Jaksch and P. Zoller, New J. Phys. {\bf 5}, 56 (2003); E. Mueller, \pra
{\bf 70}, 041603(R) (2004); K. Osterloh, M. Baig, L. Santos,
P. Zoller, and M. Lewenstein, \prl {\bf 95}, 010403
(2005); N. Goldman, A. Kubasiak, P. Gaspard, and M. Lewenstein,
\pra {\bf 79}, 023624 (2009); I. B. Spielman, {\it ibid.}
{\bf 79}, 063613 (2009).

\bibitem{nonabph1} V. Galitski, and I. B. Spielman, {\it Nature}
{\bf 494}, 49 (2013); N. Goldman, G. Juzeli\=unas, P. \"Ohberg and I B Spielman, Rep. Prog. Phys. {\bf
77}, 126401 (2014); T. Grass, K. Saha, K. Sengupta, and M.
Lewenstein, Phys. Rev. A {\bf 84}, 053632 (2011).

\bibitem{sorefs1}  Y.-J. Lin, K. Jim\'enez-García and I. B. Spielman, {\it Nature} {\bf 471}, 83-86 (2011).

\bibitem{sorefs2} Y. Li, G. I. Martone and S. Stringari, Annual Review of Cold Atoms
and Molecules, Vol. 3 (World Scientific, Singapore, 2015), Chap. 5, pp. 201-250;
Y. Li, L. P. Pitaevskii, and S. Stringari, \prl {\bf 108}, 225301 (2012).


\bibitem{sorefs3} J. Radic, A. di Colo, K. Sun, and V. Galitski, Phys. Rev.
Lett. {\bf 109}, 085303 (2012); W. S. Cole, S. Zhang, A.
Paramekanti, and N. Trivedi, Phys. Rev. Lett. {\bf 109}, 085302
(2012);

\bibitem{sorefs4} S. Mondal, K. Saha, and K. Sengupta, \prb {\bf 86}, 155101
(2012); S. Sinha, R. Nath, and L. Santos, Phys. Rev. Lett. {\bf
107}, 270401 (2011); Z. Cai, X. Zhou, and C. Wu, Phys. Rev. A {\bf
85}, 061605(R) (2012).

\bibitem{aaso} L. Zhou, H. Pu, and W. Zhang, \pra {\bf 87}, 023625 (2013).

\bibitem{Fisher} M. E. Fisher, M. N. Barber and D. Jasnow, \pra {\bf 8}, 1111 (1973).

\bibitem{supp1} See supplementary materials for more details.

\bibitem{legg1} Anthony J. Leggett, \rmp {\bf 73}, 307 (2001).

\bibitem{XXZ} T. D. Kuhner, S. R. White, and H. Monien, \prb {\bf 61}, 12474 (2000).

\bibitem{SPati} B. Pandey, S. Sinha and S. K. Pati \prb {\bf 91}, 214432 (2015).

\bibitem{Haake_p} G. Lenz and F. Haake, \prl {\bf 65}, 2325 (1990); G. Lenz and F. Haake, \prl {\bf 67}, 1 (1991); S. Schierenberg, F. Bruckmann and T. Wettig, \pre {\bf 85}, 061130 (2012).

\bibitem{evec_dist} M V Berry and M Robnik, J. Phys. A: Math. Gen. {\bf 17}, 2413 (1984); K. \`Z yczkowski, M. Lewenstein, M. Ku\'s and F. Izrailev, \pra {\bf 45}, 811 (1992).

\bibitem{Bogomolny} Y. Y. Atas, E. Bogomolny, O. Giraud, and G. Roux, \prl {\bf 110}, 084101 (2013).

\bibitem{Haake_b} F. Haake, {\it Quantum Signatures of Chaos}, Springer Science and Business Media (Springer, Berlin, Heidelberg, 2013), Vol. 54.

\bibitem{Izrailev_rmt} F. M. Izrailev, Phys. Rep. {\bf 196}, 299 (1990); V. Zelevinsky, B. A. Brown,  N. Frazier and M. Horoi, Phys. Rep. {\bf 276}, 85 (1996).

\bibitem{Varga} J. Pipek and I. Varga, \pra {\bf 46}, 3148 (1992); J. Jacquod and I. Varga, \prl {\bf 89}, 134101 (2002).

\bibitem{exp3} J. Stenger, S. Inouye, D. M. Stamper-Kurn, H. J. Miesner, A. P.
Chikkatur, and W. Ketterle, Nature (London) {\bf 396}, 345 (1998).

\bibitem{greiner1} J. Simon, W. S. Bakr, R. Ma, M. E. Tai, P. M. Preiss, and M. Greiner, Nature {\bf 472}, 307(2011)

\end{thebibliography}

\begin{thebibliography}{99}

\bibitem{stringari} Y. Zhang, Z. Yu, T. K. Ng, S. Zhang, L. Pitaevskii and S. Stringari, \pra {\bf 94} 033635 (2016).

\bibitem{shlyapnikov1d} D. S. Petrov, D. M. Gangardt and G. V. Shlyapnikov, J. Phys. IV France {\bf 116}, 3-44 (2004).

\bibitem{stringari1} Y. Li, L. P. Pitaevskii, and S. Stringari, \prl {\bf 108}, 225301 (2012)

\end{thebibliography}
\end{document}